\documentclass[useAMS,usenatbib]{mnras}
\usepackage{graphicx}
\usepackage{multirow, bigstrut}%
\usepackage{epstopdf}%
\usepackage{color}%
\usepackage{soul}%
\usepackage{amssymb}
\usepackage{amsmath}
\usepackage[usenames,dvipsnames]{xcolor}%
\renewcommand{\vec}[1]{\mathbf{#1}}
\begin{document}
\title[Isothermal shock-filament interaction]{The isothermal evolution of a shock-filament interaction}
\author[K. J. A. Goldsmith and J. M. Pittard] 
  {K. J. A. ~Goldsmith\thanks{pykjag@leeds.ac.uk} and J. M. ~Pittard\\
   School of Physics and Astronomy, University of Leeds, 
   Woodhouse Lane, Leeds LS2 9JT, UK}
   
\date{Accepted ... Received ...; in original form ...}

\pagerange{\pageref{firstpage}--\pageref{lastpage}} \pubyear{2019}
\newtheorem{theorem}{Theorem}[section]
\label{firstpage}

\maketitle

\begin{abstract}
Studies of filamentary structures that are prevalent throughout the interstellar medium are of great significance to a number of astrophysical fields. Here, we present 3D hydrodynamic simulations of shock-filament interactions where the equation of state has been softened to become almost isothermal. We investigate the effect of such an isothermal regime on the interaction (where both the shock and filament are isothermal), and we examine how the nature of the interaction changes when the orientation of the filament, the shock Mach number, and the filament density contrast are varied. We find that only sideways-oriented filaments with a density contrast of $10^{2}$ form a three-rolled structure, dissimilar to the results of a previous study. Moreover, the angle of orientation of the filament plays a large role in the evolution of the filament morphology: the greater the angle of orientation, the longer and less turbulent the wake. Turbulent stripping of filament material leading to fragmentation of the core occurs in most filaments; however, filaments orientated at an angle of $85^{\circ}$ to the shock front do not fragment and are longer-lived. In addition, values of the drag time are influenced by the filament length, with longer filaments being accelerated faster than shorter ones. Furthermore, filaments in an isothermal regime exhibit faster acceleration than those struck by an adiabatic shock. Finally, we find that the drag and mixing times of the filament increase as the angle of orientation of the filament is increased.

\end{abstract}

\section{Introduction} 
Filamentary structures are found ubiquitously in the interstellar medium (ISM). Some filaments are known to collapse under gravity and fragment into star-forming cores. Recent observational studies have further shown that a large number of prestellar cores are found within dense filaments \citep{Andre10, Arzoumanian11, Roy15}, with some filaments showing several cores strung out along their length \citep{Schisano14, Konyves15}. In addition, young stellar clusters appear at the intersections of these filaments \citep{Myers11, Schneider12}. 

The presence of cores embedded within filaments suggests a relationship between their formation and the fragmentation of the filaments themselves \citep{Schneider79, Larson85}. The conditions under which filaments fragment have been probed by various numerical studies \citep[see e.g.][]{Heigl16}. \citet{Mellema02} and \cite{Fragile04} found that radiative clouds tended to break up into clumps rather than become mixed into the background medium, since radiative or isothermal regimes can lead to milder cloud destruction. Indeed, radiative cooling can be rapid enough that almost all of the cloud mass is compressed into these long-lived clumps which would be likely to go on to collapse and form stars.

Isothermal filaments, and particularly their fragmentation, have been well studied on a theoretical basis in previous years. However, such studies have tended to assume an isothermal filament of infinite length \citep[i.e. an isothermal cylinder, as proposed by][]{Ostriker64}, which is unrealistic \citep[see][]{Chira18}. The assumption of isothermality on its own, however, may be reasonable under certain circumstances \citep[see e.g.][]{Heigl16}, though some studies have found that observed filament properties are better described by even softer equations of state \citep[e.g.][]{Toci15, Hosseinirad18, Dicintio18}. 

Our previous papers, \citet{Pittard16a} and \citet{Goldsmith16}, investigated the hydrodynamic and magnetohydrodynamic adiabatic interaction between a shock and a filament. Here, we extend the hydrodynamic study into the isothermal regime in order to understand the effects of strong radiative losses on the interaction. In the current study, all calculations are performed for a quasi-isothermal gas ($\gamma = 1.01$). Our calculations are scale-free and are applicable to a broad range of scenarios.

The outline of this paper is as follows: in Section 2 we define the isothermal shock-filament problem and review the relevant literature. Section 3 introduces the numerical method and describes the initial conditions, whilst in Section 4 we present our results. Section 5 provides a summary of our results and a conclusion. A resolution study is presented in Appendix~\ref{sec:restest}.

\section{Problem definition}
In this study, we consider the most basic scenario of a shock striking a filament. The simulated cloud is an idealised non-magnetised filament comprising a central cylindrical core of length $l r_{\rm c}$ (where $r_{\rm c}$ is the filament radius) and hemispherical caps at each end. Thus, a cloud with $l=0$ would be a spherical cloud. The total length of the filament is given by ($l + 2$)$r_{\rm c}$, and the ratio of the lengths of the major and minor axes is given by $(l+2)/2$. We vary the aspect ratio and orientation (denoted by the angle, $\theta$\footnote{$\theta = 0^{\circ}$ presents the filament as sideways-on to the shock, and $\theta = 90^{\circ}$ presents the filament as end-on to the shock.}, between the leading surface of the shock and the filament's major axis) of the filament in order to investigate how such changes might alter the interaction. Although this is clearly an idealised set-up, it is suitable for our purposes and allows changes to the interaction to be monitored as the shock Mach number, $M$, cloud density contrast, $\chi$, $l$, and $\theta$ are varied. 

The filament is initially in pressure equilibrium with its surroundings and is assumed to have smooth edges over about 10 per cent of its radius. We adopt the density profile given in \citet{Pittard09} with $p_{1}=10$, in line with our previous hydrodynamical (HD) shock-filament study, \citet{Pittard16a}. The presence of a soft edge to the filament is expected to retard the formation of HD instabilities \citep[see e.g.][]{Nakamura06, Pittard16b}. HD instabilities are expected to be further suppressed by the use of a quasi-isothermal equation of state. 

This work comprises a purely hydrodynamic study, ignoring the effects of thermal conduction, cooling, magnetic fields, and self-gravity. All calculations are performed using a softened equation of state ($\gamma =1.01$) in order to approximate an isothermal interaction (i.e. both the shock and the filament are isothermal).

\subsection{Previous work}
Numerical studies investigating the idealised problem of an adiabatic shock or wind interacting with a cloud date back to the 1970s. Since then, more realistic scenarios involving thermal conduction \citep[e.g.][]{Orlando08}, turbulence \citep[e.g.][]{Pittard09, Pittard10, Pittard16b}, and magnetic fields \citep[e.g.][]{MacLow94, Shin08}, have been published. In particular, numerical studies of shock-cloud interactions which have included radiative cooling routines include \citet{Mellema02, Fragile04, Fragile05, Orlando05, Cooper08, Cooper09, Yirak10, vanLoo10, Li13}, and \citet{Johansson13}.

Whilst a full radiative cooling model would provide more realistic results that are directly applicable to observations, such models lose their generality by introducing a physical scale (the cooling length) into the problem. Instead, softening the equation of state so that it is isothermal ($\gamma = 1$) or quasi-isothermal mimics the effects of strong cooling in the ISM whilst keeping the calculations scale-free. Several studies have explored the effect of a softened equation of state on the interaction between a shock or wind and a cloud. \citet{Klein94} briefly explored a 2D shock-cloud interaction using $\gamma=1.1$ (for the cloud only) and found that a softened equation of state led to greater compression of the cloud and reduced drag. In addition, such clouds survived for longer due to the higher density of the shocked cloud. \citet{Li03} explored self-gravitating turbulent clouds over a range of values for $\gamma$ and found that the ability of interstellar gas clouds to fragment under the action of self-gravity decreased with increasing $\gamma$ in the range $0.2 <\gamma < 1.4$. \citet{Larson05}, in a review paper, noted that the above result had particular importance for filamentary clouds, and that a value of $\gamma =1$ denoted a critical value for filament collapse. \citet{Nakamura06}, in their 3D study in which they compared adiabatic and isothermal interactions, used $\gamma=1.1$ for both the cloud and the intercloud gas. Their results supported those of \citet{Klein94} and underlined the fact that the cloud experienced much milder destruction by HD instabilities. Finally, \citet{Banda16, Banda18, Banda19} briefly explored the effect of a quasi-isothermal equation of state ($\gamma = 1.1$) on a magnetohydrodynamic (MHD) wind-cloud interaction and noted that an isothermal filament survived for longer than an adiabatic one. Other studies to have utilised an isothermal equation of state for the cloud include \citet{Raga05} and \citet{Pittard05}.

Although there is now a comprehensive, and growing, body of work concerning shock-cloud and wind-cloud interactions using spherical clouds (see \citealt{Pittard10, Pittard16b, Goldsmith17a} and \citealt{Banda19} for a brief overview), there remains a paucity of work in the current literature concerning interactions where the cloud is non-spherical. HD simulations with prolate clouds were performed by \citet{Klein94} and \citet{Xu95}, whilst \citet{Pittard16a} investigated idealised filaments. In addition, \citet{Goldsmith16} studied shock-filament interactions in a magnetised medium. With the exception of \citet{Klein94}, these studies emphasised the influence on the interaction of the alignment or orientation of the cloud with respect to the shock normal. Furthermore, \citet{Goldsmith16} noted that the length of the filament was not as important to the interaction as the filament's orientation. Other studies \citep[e.g.][]{Cooper09, Banda16, Banda18} that have investigated the formation and evolution of filamentary clouds have started from the basis of a spherical cloud (though \citet{Cooper09} also simulated a fractal cloud).

To our knowledge, there exists no other numerical study of a shock striking a filament in a non-magnetised medium using a softened equation of state. The current work, therefore, extends the study by \citet{Pittard16a} into the quasi-isothermal regime.

\subsection{Astrophysical context}
There are many situations where shock-filament interactions occur. These include cases such as a molecular cloud/filament hit by a SNR \citep*[e.g.][and references therein]{Jiang10,Vink12,Zhou14,Slane15} or by a passing shock in the ISM, a smaller filament in a molecular cloud hit by a shock resulting from nearby stellar feedback (e.g. a shock driven by an ionization front, a shock driven by a wind-blown bubble, or a shock resulting from a supernova explosion \citep[e.g.][]{McEwen16}), high-velocity clouds interacting with galaxy haloes \citep[e.g.][]{Putman12,Gronnow17}, and a molecular cloud in a multi-phase galactic wind that is interacting with faster outflowing material driven either by a stellar or active galactic nucleus activity \citep*[e.g.][]{Veilleux05,Michiyama18}. In each of these cases the interaction may occur across a wide range of parameter space. For instance, a range of Mach numbers and density contrasts could occur in each situation.

Another issue is that in many of these interactions there will be physics that is not scale-free (e.g. cooling, gravity, etc.). For instance, in the case of a filamentary IRDC that is hit by a shock in the ISM, the cloud is massive and gravity may significantly affect how material is removed from the cloud, perhaps even causing some material to fall back onto the cloud once the shock has passsed. Thus, future work which includes additional scale-dependent physics is still needed.

\section{The numerical setup}
The computations in this study were performed on a 3D $xyz$ Cartesian grid using the \textsc{mg} adaptive mesh refinement (AMR) HD code. \textsc{mg} solves numerically the Eulerian equations of hydrodynamics for the conservation of mass, momentum, and energy,
\begin{equation}
\frac{\partial \rho}{\partial t} + \nabla \cdot (\rho \vec{u}) = 0,
\end{equation}

\begin{equation}
\frac{\partial \rho \vec{u}}{\partial t} + \nabla \cdot (\rho \vec{u} \vec{u}) + \nabla P = 0 ,
\end{equation}

\begin{equation}
\frac{\partial E}{\partial t} + \nabla \cdot [(E+P) \vec{u}] = 0,
\end{equation}
respectively, where $\rho$ is the mass density, $\vec{u}$ is the velocity, $P$ is the thermal pressure, $\gamma$ is the ratio of specific heat capacities, and

\begin{equation}
E=\frac{P}{\gamma -1} + \frac{1}{2} \rho \vec{u}^{2}
\end{equation}
is the total energy density.

\textsc{mg} uses piecewise linear cell interpolation and solves the Riemann problem at each cell interface in order to determine the conserved fluxes for the time update. The scheme is second-order accurate in space and time. A linear solver is used in most instances, with the code switching to an exact solver where there is a large difference between the two states \citep{Falle91}. 

The two coarsest levels ($G^{0}$ and $G^{1}$) of the AMR grid cover the entire computational domain, with finer grids being added where needed and removed where they are not. Refinement and derefinement are performed on a cell-by-cell basis and are controlled by the differences in the solutions on the coarser grids at any point in space (see \citealt{Goldsmith17a} for a more detailed description of the refinement process). Grid level $G^{0}$ has a cell width of $4 \,r_{c}$. The effective spatial resolution of each simulation is taken to be the resolution of the finest grid and is given by $R_{\rm cr}$, where `cr' is the number of cells per filament semi-minor axis in the finest grid, equivalent to the number of cells per cloud radius for a spherical cloud. This effective radius is taken to be the `filament radius'. Each of the simulations was performed at an effective resolution of $R_{16}$, requiring seven grid levels ($G^{0}$ to $G^{6}$). A resolution study is presented in Appendix~\ref{sec:restest}.

The filament is initially centred at the grid origin $(x, y, z) = (0, 0, 0)$ with the planar shock front located at $x = -10$. The shock propagates along the $x$-axis in the positive $x$ direction. The numerical domain is set so that there is constant inflow from the negative $x$ direction and free inflow/outflow conditions at other boundaries, and is large enough so that the main features of the interaction occur before the shock reaches the downstream boundary of the grid. The grid extent is determined by the values of $M$, $\chi$, and $\theta$ and is $-20 < x < 1500$, $-12 < y < 12$, $-12 < z < 12$ for all simulations. In this study, we define motion in the direction of shock propagation as `axial' and that perpendicular to this as `radial' or `transverse' (this includes motion in both the $y$ and $z$ directions). 

All length scales are measured in units of the filament radius, $r_{\rm c}$, where $r_{\rm c}=1$, velocities are measured in units of the shock velocity through the ambient medium, $v_{b}$, and the unit of density is given as the density of the surrounding pre-shocked gas, $\rho_{\rm amb}$. For a Mach 3 shock with $\gamma = 1.01$ the post-shock density, pressure, and velocity relative to the pre-shock ambient values and to the shock speed are (in computational units) $\rho_{\rm ps}/\rho_{\rm amb} = 8.7$, $P_{\rm ps}/P_{\rm amb} = 9.1$, and $v_{\rm ps}/v_{b} = 0.88$, respectively.

\subsection{Diagnostics}
Various integrated quantities allow the evolution of the filament to be studied \citep[see][]{Klein94, Nakamura06, Pittard09, Goldsmith16}. Averaged quantities, $\langle f \rangle$, are constructed by
\begin{equation}
\langle f \rangle = \frac{1}{m_{\beta}} \int_{\kappa \geq \beta} \kappa \rho f \; \mathrm{d}V ,
\end{equation}
where $m_{\beta}$, the mass identified as being part of the filament, is given by
\begin{equation}
m_{\beta} = \int_{\kappa \geq \beta} \kappa \rho \; \mathrm{d}V .
\end{equation}
An advected scalar, $\kappa$, is used to trace the filament material in the flow, allowing the whole filament along with its denser core to be distinguished from the ambient medium. {\bf $\kappa = \rho/(\chi \rho_{\rm amb})$}. It has an initial value of 1.0 at the centre of the filament and declines towards the filament edge, reducing to a value of zero for the surrounding ambient material. $\beta$ is the threshold value, and integrations are only performed over cells where $\kappa \geq \beta$. Setting $\beta=0.5$ allows the densest regions of the filament and its associated fragments (hereafter subscripted as `core') to be probed, whilst setting $\beta=2/\chi$ probes the entire filament and its low-density envelope, as well as regions where some filament material is mixed into the ambient flow (hereafter subscripted as `cloud'). Note that our meaning of the term `core' differs from the `core'-like structures that are understood by the star formation community. In our work it simply refers to dense fragments.

The mass-weighted mean velocity of the filament in each direction ($\langle v_{x}\rangle$, $\langle v_{y}\rangle$, $\langle v_{z}\rangle$) and the velocity dispersions in all three directions, defined as
\begin{equation}
\delta v_{x} = \left( \langle v^{2}_{x} \rangle - \langle v_{x} \rangle^{2} \right)^{1/2},
\end{equation}
\begin{equation}
\delta v_{y} = \left( \langle v^{2}_{y} \rangle - \langle v_{y} \rangle^{2} \right)^{1/2},
\end{equation}
\begin{equation}
\delta v_{z} = \left( \langle v^{2}_{z} \rangle - \langle v_{z} \rangle^{2} \right)^{1/2},
\end{equation}
are followed, as well as the mean density, which is given as
\begin{equation}
\langle \rho \rangle = \frac{m_{\beta}}{V_{\beta}},
\end{equation}
where $V_{\beta}$ is the volume of a region having $\kappa \geq \beta$.

\subsection{Dynamical time-scales}\label{time}
Time zero in our calculations is taken to be the time at which the shock is level with the centre of the filament. The characteristic time-scale for a spherical cloud to be crushed by the shocks being driven into it is the so-called `cloud-crushing time' defined by \citet{Klein94}. However, a modified time-scale for prolate clouds to be crushed by the shock being driven into them was adopted by \citet{Xu95},
\begin{equation}
t_{cs} = \frac{r_{\rm s}\chi^{1/2}}{v_{b}}\, ,
\end{equation}
where $r_{\rm s}$ is the radius of a spherical cloud of equivalent mass, and is used throughout this paper, in line with \citet{Pittard16a}.

Several other time-scales can be obtained. The time taken for the average filament velocity relative to that of the post-shock flow, $v_{\rm ps}$ (as measured in the frame of the pre-shock ambient medium), to decrease by a factor of $e$ (i.e. the time when the average filament velocity $\langle v \rangle_{\rm cloud}\,=(1-1/e)\,v_{\rm ps}$) is known as the `drag time', $t_{\rm drag}$; the `mixing time', $t_{\rm mix}$, is the time at which the filament core mass, $m_{\rm core}$, has reached half that of its initial value; and the filament `lifetime', $t_{\rm life}$, is defined as the time when the filament core mass has reached one per cent of its initial value.
    
\section{Results}
In this section we begin by examining the morphology of the interaction for our reference simulation, model $m3c2l8s$, and then consider the morphology for simulations with $M=3$ for clouds of varying length and orientation, comparing against calculations made using a shock of $M=10$ in an adiabatic regime. At the end of this section, we explore the impact of the interaction on various global quantities and time-scales. Table~\ref{Table1} summarises the calculations performed and provides some key time-scales, whilst Table~\ref{Table2} provides the shock jump values for simulations with $M=1.5, \, 3,$ and 10. We adopt a naming convention such that $m3$ denotes $M=3$, $c2$ denotes $\chi=10^{2}$, $l8$ denotes a filament length of 8, and $s$ refers to a filament orientated sideways to the shock front (sideways filaments have $\theta=0^{\circ}$; where the orientation of the filament is other than sideways, the number given in the model name refers to the angle of orientation of the major axis to the shock front). 

Compared to our previous work in \citet{Pittard16a}, the softer equation of state causes significantly greater compression of the filament. Thus, the cross-section of the filament is smaller than that of the adiabatic filament (with $\gamma =5/3$). Whilst this effect tends to reduce the efficiency of the ambient flow in dragging the cloud with it, it is countered by the higher post-shock intercloud density (which in the $\gamma=1.01$ case is 2.89 times higher for $M=3$ and 17.3 times higher for $M=10$, relative to the $\gamma =5/3$ case). This higher intercloud density, plus the faster post-shock speed, gives a greater drag force. We find that this latter effect is dominant, so that the drag times are much shorter for the isothermal ($\gamma =1.01$) filaments compared to the adiabatic ($\gamma =5/3$) filaments. A key consequence of this greater acceleration is that the relative velocity of the ambient and cloud material rapidly declines, such that Kelvin-Helmholtz (KH) and Rayleigh-Taylor (RT) instabilities are mild (see also \citealt{Nakamura06}).

\begin{table*} \footnotesize
\setlength{\tabcolsep}{5pt}
\centering
 \caption{A summary of the shock-filament simulations presented in this work, along with key time-scales. $M$ is the shock Mach number, $\chi$ is the density contrast of the filament to the surrounding ambient medium, $l$ defines the length of the filament, and $\theta$ defines the angle of orientation of the filament between its major-axis and the shock surface (sideways-oriented filaments have $\theta = 0^{\circ}$). $v_{\rm ps}$ is the post-shock flow velocity, and is given in units of $v_{b}$. $t_{\rm cc}$ is the cloud-crushing time-scale of \citet{Klein94}, while $t_{\rm cs}$ is the cloud-crushing timescale for a spherical cloud of equivalent mass introduced by \citet{Xu95}. Key filament time-scales are additionally noted. $\dag$ denotes that the true value is unable to be given because the simulation had ended before this point was reached. Note that simulations $m3c2l885$, $m3c3l8s$, and $m10c2l885$ were run at a reduced resolution of $R_{16}$.}
 \label{Table1}
 \begin{tabular}{@{}lccccccccll}
  \hline
  Simulation & $M$ & $\chi$ 
        & $l$ ($r_{\rm c}$) & $\theta$ ($^{\circ}$) & $v_{\rm ps}/v_{b}$ &  $t_{\rm cs}/t_{\rm cc}$ & $t_{\rm drag}/t_{\rm cs}$ & $t_{\rm mix}/t_{\rm cs}$ & $t_{\rm life}/t_{\rm cs}$ \\
  \hline
m10c1l8s & 	10 & 	$10$ & 8 & sideways  & 0.99 & 1.91 & 0.61 &0.44  & $-^{\dag}$ \\
m10c2l2s & 	10 & 	$10^{2}$ & 2 & sideways & 0.99 & 1.36 & 0.43 &0.27 & 0.50 \\
m10c2l4s & 	10 & 	$10^{2}$ & 4 & sideways & 0.99 & 1.59 & 0.36 & 0.23 & 0.43 \\
m10c2l8s &      10 & 	$10^{2}$ & 8 & sideways &  0.99 & 1.91 & 0.30 & 0.19 & 0.36 \\
m10c2l230 & 	10 & 	$10^{2}$ & 2 & 	$30^{\circ}$ &  0.99 & 1.35 & 0.51 & 0.27 & 0.66 \\
m10c2l430 & 	10 & 	$10^{2}$ & 4 & 	$30^{\circ}$ & 0.99 & 1.58 & 0.43 &0.23  & 0.57 \\
m10c2l830 & 	10 & 	$10^{2}$ & 8 & 	$30^{\circ}$ &  0.99 & 1.91 &0.36 & 0.19 & 0.47 \\
m10c2l860 & 	10 & 	$10^{2}$ & 8 & 	$60^{\circ}$ &  0.99 & 1.91 & 0.71 & 0.41 & 1.54 \\
m10c2l885 & 	10 & 	$10^{2}$ & 8 & 	$85^{\circ}$ &  0.99 & 1.91 & 1.53 & 0.98 & 4.93 \\
m3c1l8s & 	3 & 	$10$ & 8 & sideways  &  0.88 & 1.91 & 0.72 & 7.06 & 23.8 \\
m3c2l2s & 	3 & 	$10^{2}$ & 2 & sideways &  0.88 & 1.36 & 2.06 &3.14 & 6.54 \\
m3c2l4s & 	3 & 	$10^{2}$ & 4 & sideways & 0.88 & 1.59 & 1.76 & 3.04 &6.27  \\
m3c2l8s & 	3 & 	$10^{2}$ & 8 & sideways &  0.88 & 1.91 & 1.46 & 2.82 &5.92  \\
m3c2l230 & 	3 & 	$10^{2}$ & 2 & 	$30^{\circ}$ &  0.88 & 1.36 &3.23  &4.14  & 7.07 \\
m3c2l430 & 	3 & 	$10^{2}$ & 4 & 	$30^{\circ}$ &  0.88 & 1.59 &  2.69& 3.61 & 6.41 \\
m3c2l830 & 	3 & 	$10^{2}$ & 8 & 	$30^{\circ}$ &  0.88 & 1.91 &2.11 & 3.12 & 5.64 \\
m3c2l860 & 	3 & 	$10^{2}$ & 8 & 	$60^{\circ}$ &  0.88 & 1.91 & 5.25 & 5.93 & 8.90 \\
m3c2l885 & 	3 & 	$10^{2}$ & 8 & 	$85^{\circ}$ &  0.88 & 1.91 &  6.40&  6.81&  10.7\\
m3c3l8s & 	3 & 	$10^{3}$ & 8 & 	sideways &  0.88 & 1.91 &  2.03 &  2.58 & 4.52 \\
m1.5c1l8s & 	1.5 & $10$ & 8 & sideways &  0.55 & 1.91 & 2.16 & 9.26 & 12.31 \\
m1.5c2l2s & 	1.5 & $10^{2}$ & 2 & sideways & 0.55 & 1.36 & 8.72 & 8.98 & 18.8 \\
m1.5c2l4s & 	1.5 & $10^{2}$ & 4 & sideways & 0.55 & 1.91 & 6.26 & 7.53 & 13.2 \\
m1.5c2l8s & 	1.5 & $10^{2}$ & 8 & sideways &  0.55 & 1.91 & 5.36 & 6.61 & 12.2 \\
m1.5c2l230 & 	1.5 & $10^{2}$ & 2 & $30^{\circ}$ &  0.55 & 1.36 & 7.73 & 9.15 & 17.96 \\
m1.5c2l430 & 	1.5 & $10^{2}$ & 4 & $30^{\circ}$ & 0.55 & 1.36 & 7.04 & 8.26 & 13.33 \\
m1.5c2l830 & 	1.5 & $10^{2}$ & 8 & $30^{\circ}$ &  0.55 & 1.36 & 6.03 & 6.85 & 13.78 \\
m1.5c2l860 & 	1.5 & $10^{2}$ & 8 & $60^{\circ}$ &  0.55 & 1.36 & 7.50 & 7.32 & 14.47 \\
m1.5c2l885 & 	1.5 & $10^{2}$ & 8 & $85^{\circ}$ & 0.55 & 1.36 & 8.08 & 7.84 & 13.66 \\
  \hline
 \end{tabular}
\end{table*}

\begin{table}
\centering
\caption{The shock jump values for $M=1.5, \, 3$, and 10.}
\label{Table2}
\begin{tabular}{@{}lc}
  \hline
 $M$ & Density/pressure jump \\
    \hline
1.5 & 2.27   \\
3 & 9.09  \\
10 & 101.0    \\
  \hline
 \end{tabular}
\end{table}

\subsection{Interaction of a filament with $\chi=10^{2}$ and a sideways orientation with a shock of $M=3$}
\begin{figure*}
\centering
\begin{tabular}{c}
\includegraphics[width=170mm]{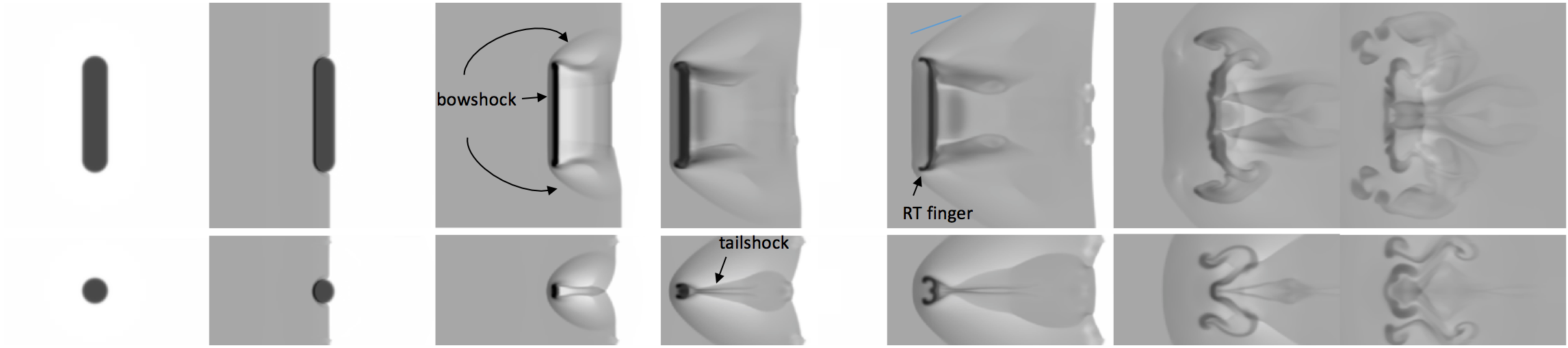} \\
\end{tabular}
\caption{The time evolution of the logarithmic density for model $m3c2l8s$ showing the $xy$ (top set of panels) and $xz$ planes (bottom set of panels). The greyscale shows the logarithm of the mass density, from white (lowest density) to black (highest density). The density in this and subsequent figures has been scaled with respect to the ambient density $\rho_{\rm amb}$, so that a value of 0 represents the value of $\rho_{\rm amb}$ and 1 represents $10 \times \rho_{\rm amb}$. The density scale used for this figure extends from 0 to 2.8. The first panel on each row shows the setup of the simulation. The evolution proceeds from the second panel onwards, left to right, with $t=0.00 \,t_{\rm cs}$, $t=0.39 \,t_{\rm cs}$, $t=0.66 \,t_{\rm cs}$, $t=0.95 \,t_{\rm cs}$, $t=2.16\,t_{\rm cs}$, and $t=2.82 \,t_{\rm cs}$. All frames show the same region for $y$ and $z$ ($-10 < y < 10$ and $-5 < z < 5$, in units of $r_{\rm c}$). So that the motion of the cloud is clear, the first 3 frames show $-10 < x < 10$. Frames 4-6 show $0 < x < 20$, and the final frame shows $20 < x < 40$. Note that in this and similar figures the $y$ and $z$ axes are plotted vertically, with positive towards the top and negative towards the bottom, whilst the shock is initially located at $x=-10$.}
\label{Fig1}
\end{figure*}

We begin by discussing the morphology of the interaction for our reference simulation, where $M=3$, $\chi=10^{2}$, $l=8$, and the filament is oriented sideways to the shock front. Figure~\ref{Fig1} shows the mass density as a function of time for the $xy$ and $xz$ planes. The first panel in this, and subsequent figures, shows the initial filament orientation, and the shock propagates from left to right. We first describe the nature of the interaction and changing morphology with reference to a filament struck by a $M=10$ adiabatic shock presented in a previous paper \citep{Pittard16a}. The rationale for focussing on a filament struck by a $M=3$ shock instead of that struck by a shock of $M=10$ is that the former is a more typical scenario for a filamentary IRDC struck by an ISM shock (which are more numerous at lowish Mach numbers).

Figure~\ref{Fig1} shows the filament being struck by the shock from its side. The second panel, at $t= 0.00 \, t_{cs}$, shows that the external shock has just passed the centre of the filament, whilst at $t= 0.39\, t_{\rm cs}$ a bow shock has formed on the upstream side of the filament, very close to its upstream edge, in contrast with the $M=10$ adiabatic simulation where the bow shock is located at a slightly greater distance from the filament (see figure 3 in \citealt{Pittard16a}). The upstream surface of the filament begins to be compressed (as evidenced by an increase in density at this point) by the transmitted shock progressing through it, while the external shock sweeps symmetrically around the outside of the filament and converges at the rear of the cloud, creating a region of higher pressure compared to the pressure of the ambient medium downstream of the cloud. The convergence of the external shock on the $z=0$ plane forces a secondary shock back through the cloud in the upstream direction.

The filament reaches maximum compression at $t\approx 0.39\, t_{\rm cs}$. At this point the transmitted shock has travelled through and exited the filament and has propagated downstream, accelerating as it proceeds and dragging filament material with it. As it exits the back of the filament the ends of the filament begin to display the effects of diffracted shocks and some ablation of filament material by the surrounding flow is observed (in line with the sideways filament in the $M=10$ adiabatic simulation presented in \citealt{Pittard16a}). The ends of the filament at this point are bent in the upstream direction; this bears some similarities with the sideways filament of length $l=4$ and density contrast $\chi=10$ embedded in a perpendicular magnetic field in \citet{Goldsmith16}, where it was noted that care ought to be taken from an observational point of view since the interpretation of such a filament might lead to the conclusion that the shock was travelling in the $-x$ direction (this effect was also visible in \citealt{Pittard16a}). After this point (at $t\approx 0.66\, t_{\rm cs}$), the filament expands due to rarefaction waves within it. A `tail shock' (as noted in \citealt{Pittard16a}) is visible. The filament is then seen to collapse in on itself at $t\approx 0.95\, t_{\rm cs}$ and become compacted in the direction of shock propagation. Small RT fingers develop on the tips of the filament. Unlike in \citet{Pittard16a}, the upstream edge of the filament displays no obvious KH instabilities at $t=0.92\, t_{\rm cs}$ due to the quasi-isothermal nature of the interaction. The filament is also much more compressed than in the aforementioned study and its tail of ablated cloud material is much smoother. As noted in \citet{Pittard16a}, the filament forms a `three-rolled' structure, though this becomes more spread out as the filament material is ablated by the flow.

\begin{figure*}
\centering
\begin{tabular}{c}
\includegraphics[width=150mm]{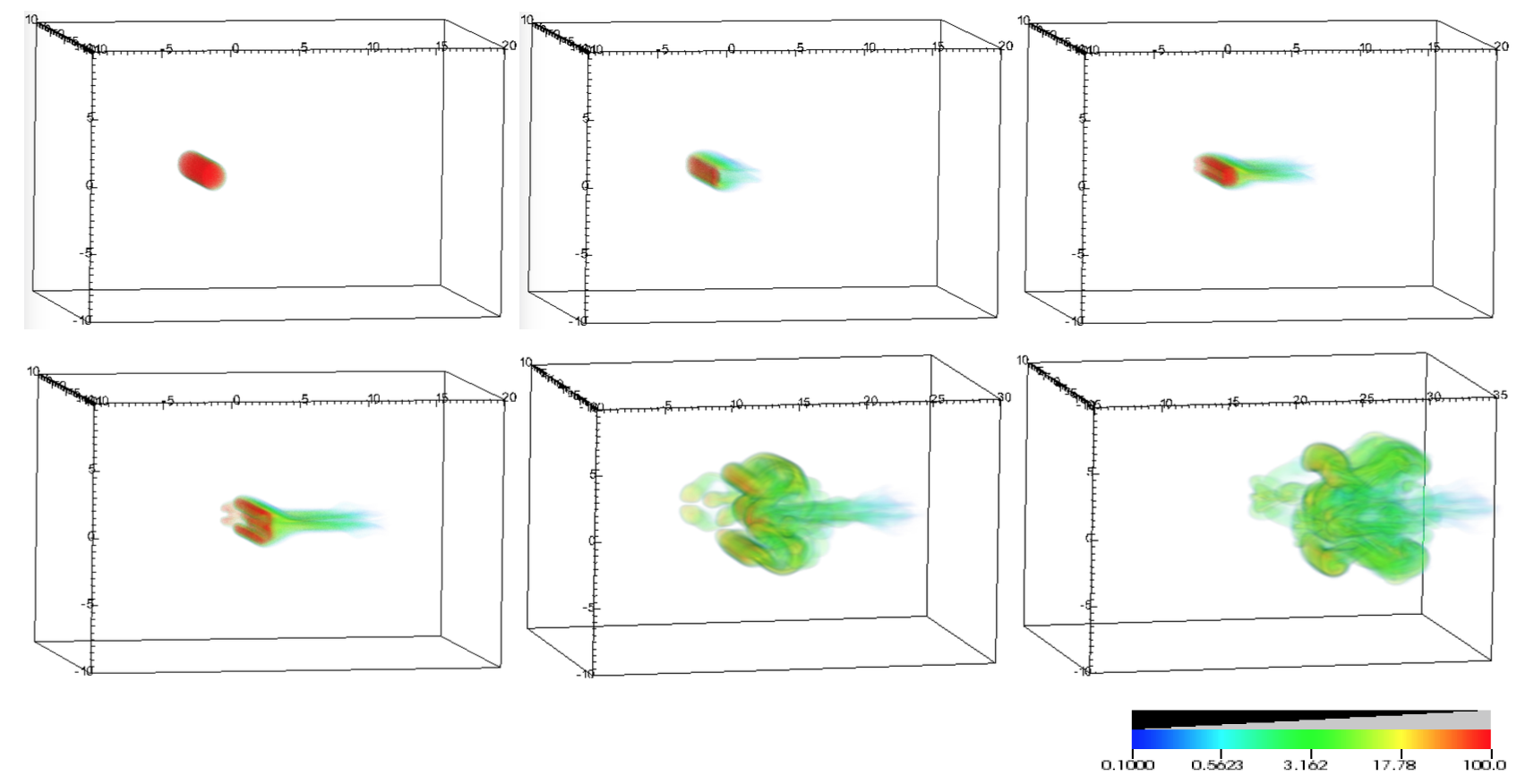} \\
\end{tabular}
\caption{A 3D volumetric rendering of model $m3c2l8s$. From left to right, top to bottom the timings are $t=0.00 \,t_{\rm cs}$, $t=0.39 \,t_{\rm cs}$, $t=0.66 \,t_{\rm cs}$, $t=0.95 \,t_{\rm cs}$, $t=2.16\,t_{\rm cs}$, and $t=2.82 \,t_{\rm cs}$. The colour scale in this and similar figures indicates the density of the filament, normalised by the density of the ambient medium, with the initial filament density being 100 (or red). The ambient medium is not shown; therefore, the bow shock upstream of the filament is also not visible.}
\label{Fig2}
\end{figure*}

Figure~\ref{Fig2} shows volumetric density renderings of the filament as a function of time in the $xz$ and $xy$ planes, respectively. Owing to the focus on filament material, this figure (and subsequent similar figures) does not show features such as the bow shock or other elements of the ambient material or flow. The main differences between this figure and figure 1 of \citet{Pittard16a} are that the entrainment of filament material by the flow is much smoother in the current figure (as expected by the lack of KH instabilities produced by the damping effect of the quasi-isothermal equation of state) and thus there is no turbulent mass of filament material located to the rear of the cloud. Moreover, a short tail of material is observed to form on the axis behind the filament as the simulation progresses. The three-rolled structure identified by \citet{Pittard16a} is present in this figure (from $t\approx 0.95\, t_{\rm cs}$ onwards).

\subsection{Effect of filament orientation on the interaction}
\begin{figure*}
\centering
\begin{tabular}{c}
\includegraphics[width=170mm]{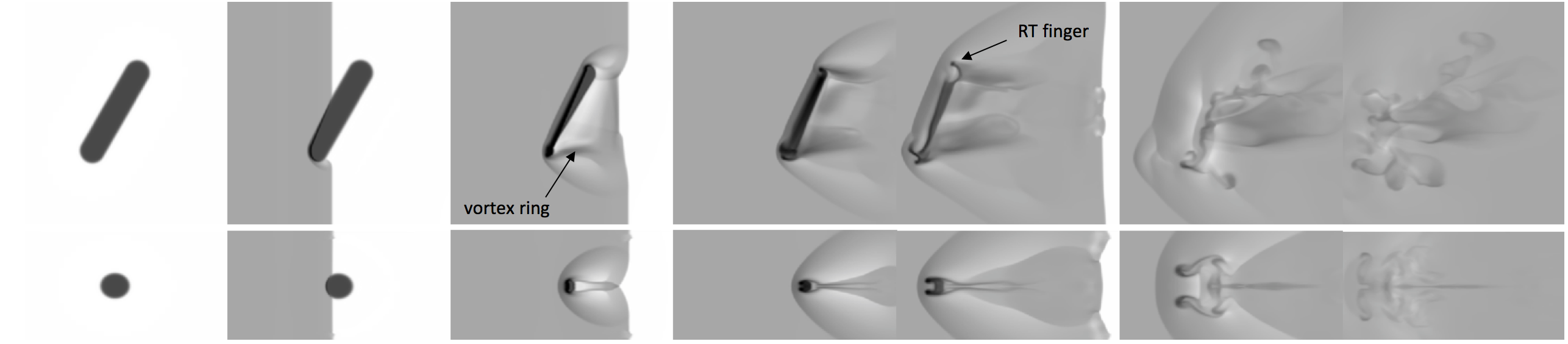} \\
\end{tabular}
\caption{The time evolution of the logarithmic density for model $m3c2l830$ showing the $xz$ (top set of panels) and $xy$ planes (bottom set of panels). The greyscale shows the logarithm of the mass density, from white (lowest density) to black (highest density). The density scale used for this figure extends from 0 to 2.8. The first panel on each row shows the setup of the simulation. The evolution proceeds from the second panel onwards, left to right, with $t=0.00 \,t_{\rm cs}$, $t=0.39 \,t_{\rm cs}$, $t=0.66 \,t_{\rm cs}$, $t=0.95 \,t_{\rm cs}$, $t=2.16\,t_{\rm cs}$, and $t=2.82 \,t_{\rm cs}$. All frames show the same region in $y$ and $z$ ($-5 < y < 5$ and $-10 < z < 10$, in units of $r_{\rm c}$). So that the motion of the cloud is clear, the first 4 frames show $-10 < x < 10$. Frame 5 shows $0 < x < 20$, frame 6 shows $5 < x < 25$, and the final frame shows $20 < x < 40$.}
\label{Fig4}
\end{figure*}

\begin{figure*}
\centering
\begin{tabular}{c}
\includegraphics[width=150mm]{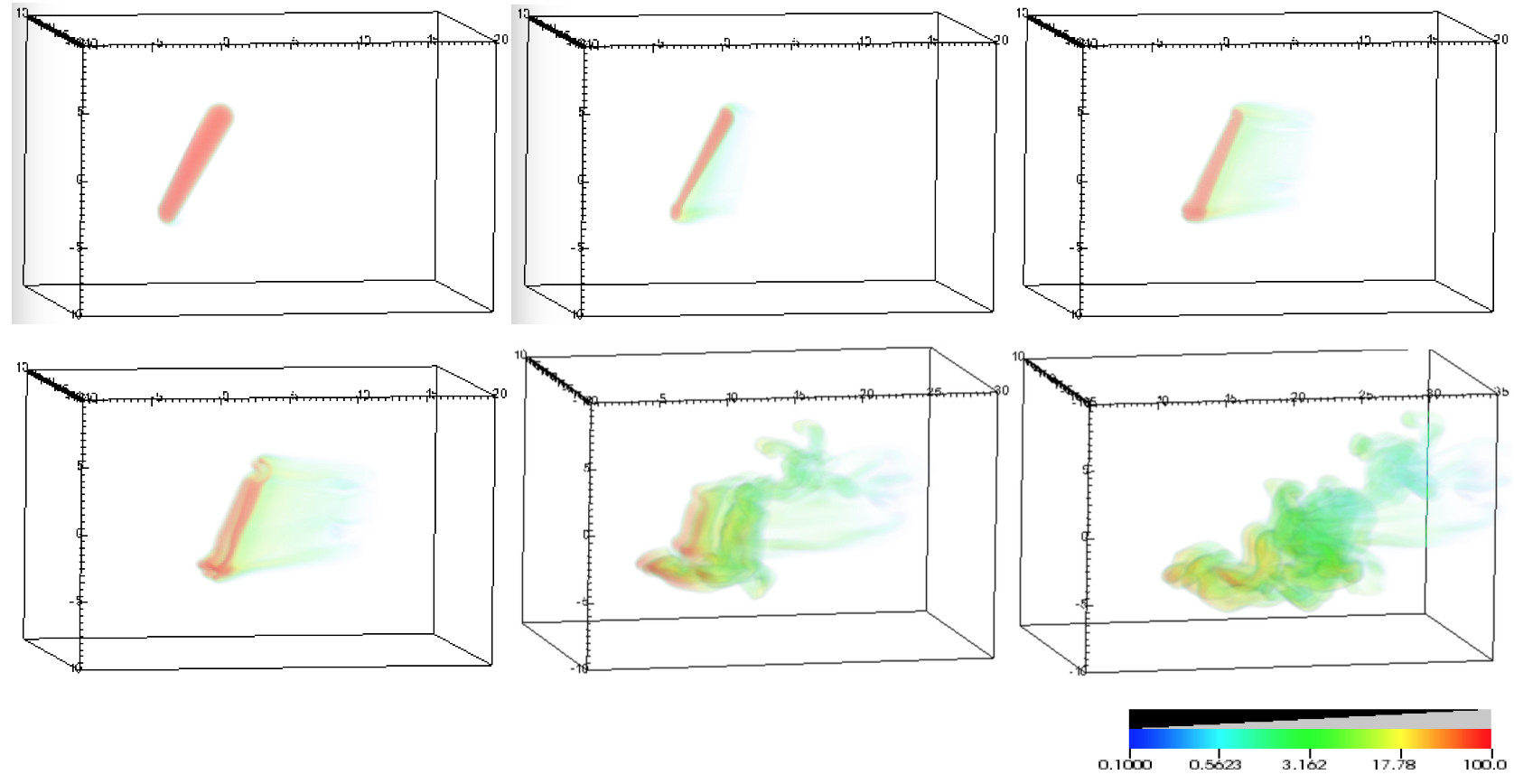} \\
\end{tabular}
\caption{A 3D volumetric rendering of model $m3c2l830$. From left to right, top to bottom the timings are $t=0.00 \,t_{\rm cs}$, $t=0.39 \,t_{\rm cs}$, $t=0.66 \,t_{\rm cs}$, $t=0.95 \,t_{\rm cs}$, $t=2.16\,t_{\rm cs}$, and $t=2.82 \,t_{\rm cs}$.}
\label{Fig5}
\end{figure*}

Figure~\ref{Fig4} shows the interaction of an $M=3$ shock with an obliquely-oriented filament (i.e. one oriented at $\theta = 30^{\circ}$ to the shock front). Unlike the sideways filament in Figs~\ref{Fig1} and ~\ref{Fig2} which was struck from the side, the filament in the current figure is initially struck at its upstream-facing end. The external shock is then channelled around the edge of the filament. By $t= 0.39\, t_{\rm cs}$ the external shock has fully diffracted around the filament and has converged on the axis behind the filament, interacting and causing shocks to be driven back into the rear of the filament at an angle. At the same time, a transmitted shock is making its way through the filament from the upstream side, leading to that part of the filament becoming compressed and the filament taking on a wedge-shaped appearance. As the shock moves through the filament, the bottom end of the filament expands whilst the top end is steadily compressed ($t\approx 0.66\, t_{\rm cs}$). Filament material begins to be ablated from each end of the filament at $t= 0.39\, t_{\rm cs}$ onwards and a vortex ring is visible at the base of the filament. Meanwhile, an RT finger is evident at the top end of the filament from $t\approx 0.95\, t_{\rm cs}$. Considerably less turbulent stripping of filament material is evident in the current figure and it is clear that the bow shock is much closer to the upstream edge of the filament compared to the bow shock in figure 7 in \citet{Pittard16a}. Because of the RT finger, the flow of filament material stripped by the surrounding flow is channelled behind and above the filament, though it is much less turbulent. At $t= 2.16\, t_{\rm cs}$ multiple shocks are present at the location of the bow shock and are caused by shocks propagating back through the filament and accelerating into the surrounding flow. Clumps of filament material are observed to break away from the top of the filament, and the filament core mass has been significantly ablated by the flow, though still retaining its structure. Figure~\ref{Fig5} shows how the filament forms a short turbulent wake at late times.
 
\begin{figure*}
\centering
\begin{tabular}{c}
\includegraphics[width=170mm]{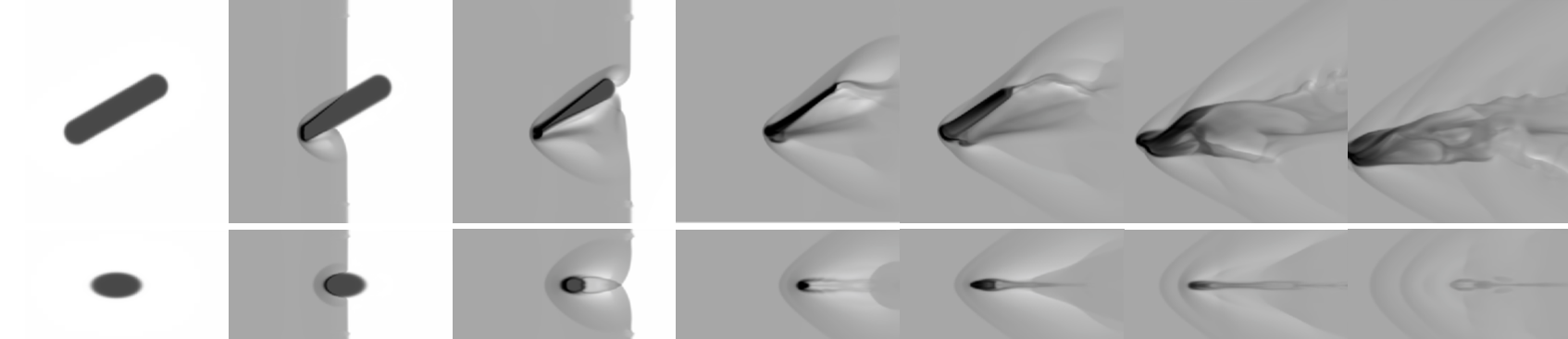} \\
\end{tabular}
\caption{The time evolution of the logarithmic density for model $m3c2l860$ showing the $xz$ (top set of panels) and $xy$ planes (bottom set of panels). The greyscale shows the logarithm of the mass density, from white (lowest density) to black (highest density). The density scale used for this figure extends from 0 to 2.8. The first panel on each row shows the setup of the simulation. The evolution proceeds from the second panel onwards, left to right, with $t=0.00 \,t_{\rm cs}$, $t=0.39 \,t_{\rm cs}$, $t=0.66 \,t_{\rm cs}$, $t=0.95 \,t_{\rm cs}$, $t=2.16\,t_{\rm cs}$, and $t=2.82 \,t_{\rm cs}$. All frames show the same region in $y$ and $z$ ($-5 < y < 5$ and $-10 < z < 10$, in units of $r_{\rm c}$). So that the motion of the cloud is clear, the first 4 frames show $-10 < x < 10$. Frame 5 shows $-5 < x < 15$, frame 6 shows $0 < x < 20$, and the final frame shows $5 < x < 25$.}
\label{Fig7}
\end{figure*}

\begin{figure*}
\centering
\begin{tabular}{c}
\includegraphics[width=150mm]{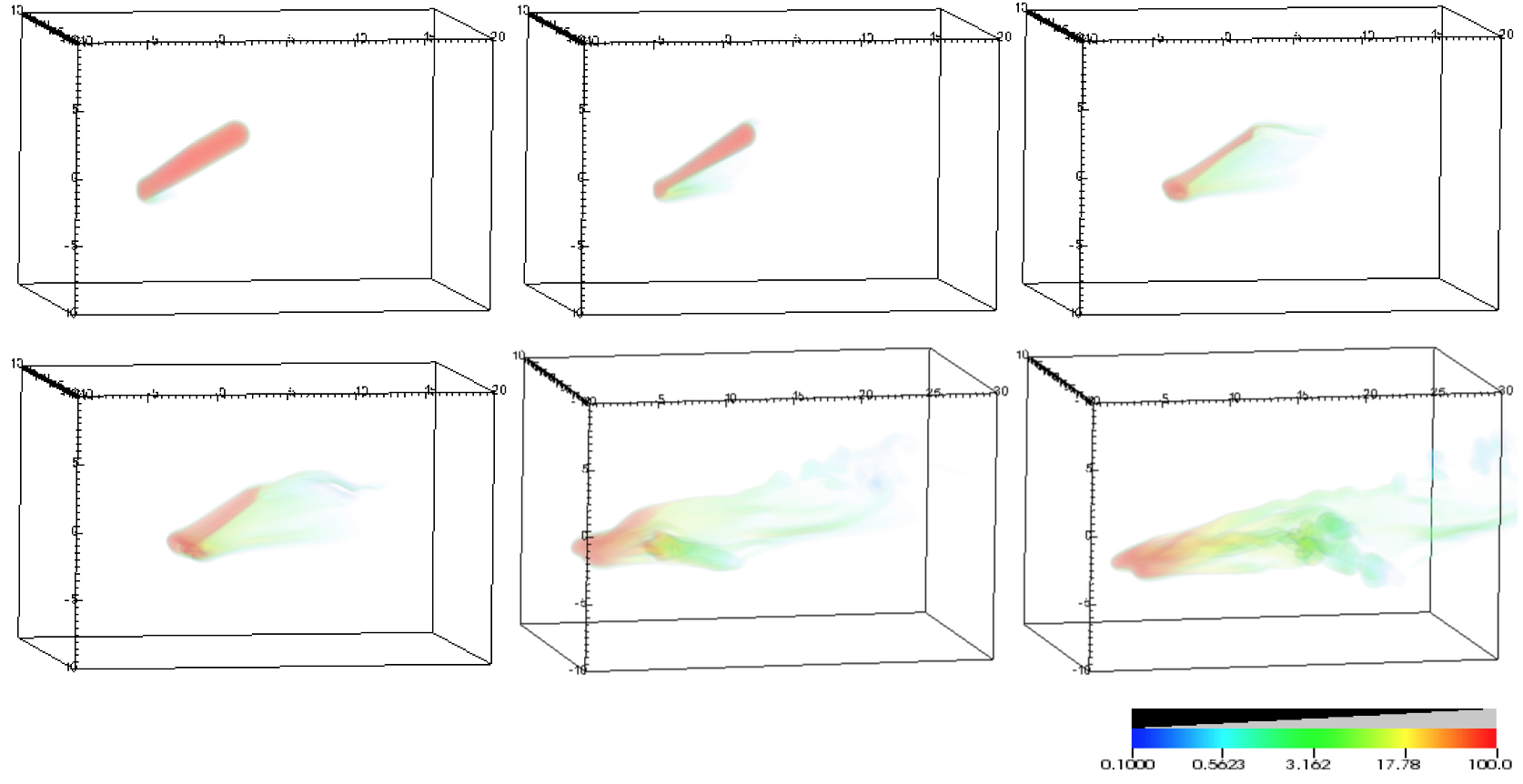} \\
\end{tabular}
\caption{A 3D volumetric rendering of model $m3c2l860$. From left to right, top to bottom the timings are $t=0.00 \,t_{\rm cs}$, $t=0.39 \,t_{\rm cs}$, $t=0.66 \,t_{\rm cs}$, $t=0.95 \,t_{\rm cs}$, $t=2.16\,t_{\rm cs}$, and $t=2.82 \,t_{\rm cs}$.}
\label{Fig8}
\end{figure*}

The interaction of a shock with a filament oriented at $\theta=60^{\circ}$ to the shock front (simulation $m3c2l860$) is shown in Figs~\ref{Fig7} and ~\ref{Fig8}. The initial morphology is not dissimilar to that in simulation $m3c2l830$. However, at later times (from $t= 0.95\, t_{\rm cs}$ onwards) the filament length becomes compressed until it is less than half its original length. The vortex ring located at the upstream end of the filament is much larger than before, whilst the RT finger at the top of the filament and its associated wake of filament material extends much further downstream. A double bow shock is observed in the $xy$ panels at later times.

\begin{figure*}
\centering
\begin{tabular}{c}
\includegraphics[width=170mm]{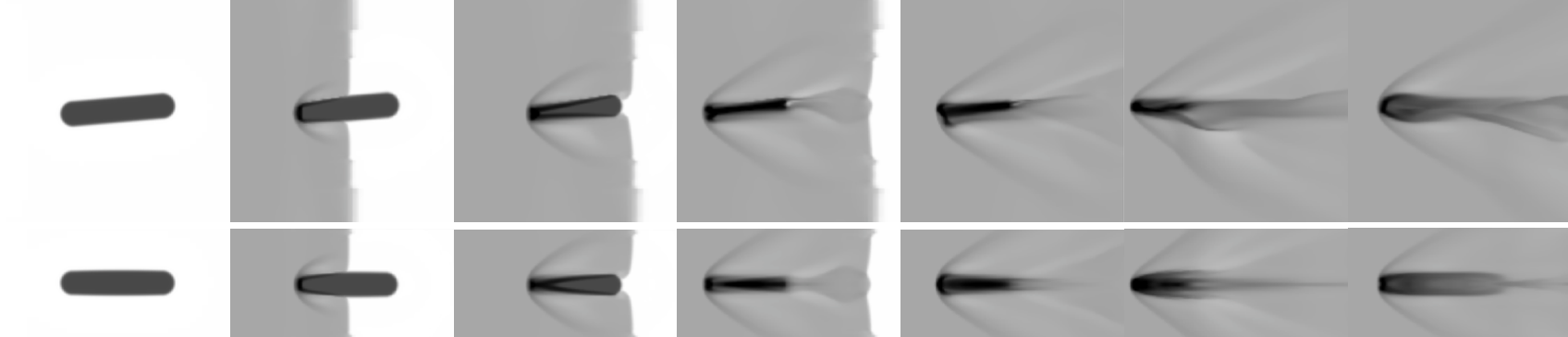} \\
\end{tabular}
\caption{The time evolution of the logarithmic density for model $m3c2l885$ showing the $xz$ (top set of panels) and $xy$ planes (bottom set of panels). The greyscale shows the logarithm of the mass density, from white (lowest density) to black (highest density). The density scale used for this figure extends from 0 to 2.8. The first panel on each row shows the setup of the simulation. The evolution proceeds from the second panel onwards, left to right, with $t=0.00 \,t_{\rm cs}$, $t=0.39 \,t_{\rm cs}$, $t=0.66 \,t_{\rm cs}$, $t=0.95 \,t_{\rm cs}$, $t=2.16\,t_{\rm cs}$, and $t=2.82 \,t_{\rm cs}$. All frames show the same region in $y$ and $z$ ($-5 < y < 5$ and $-10 < z < 10$, in units of $r_{\rm c}$). So that the motion of the cloud is clear, the first 3 frames show $-10 < x < 10$. Frames 4 and 5 show $-5 < x < 15$, and the final two frames show $0 < x < 20$. Note that this simulation was run at a slightly lower resolution of $R_{16}$.}
\label{Fig10}
\end{figure*}

\begin{figure*}
\centering
\begin{tabular}{c}
\includegraphics[width=150mm]{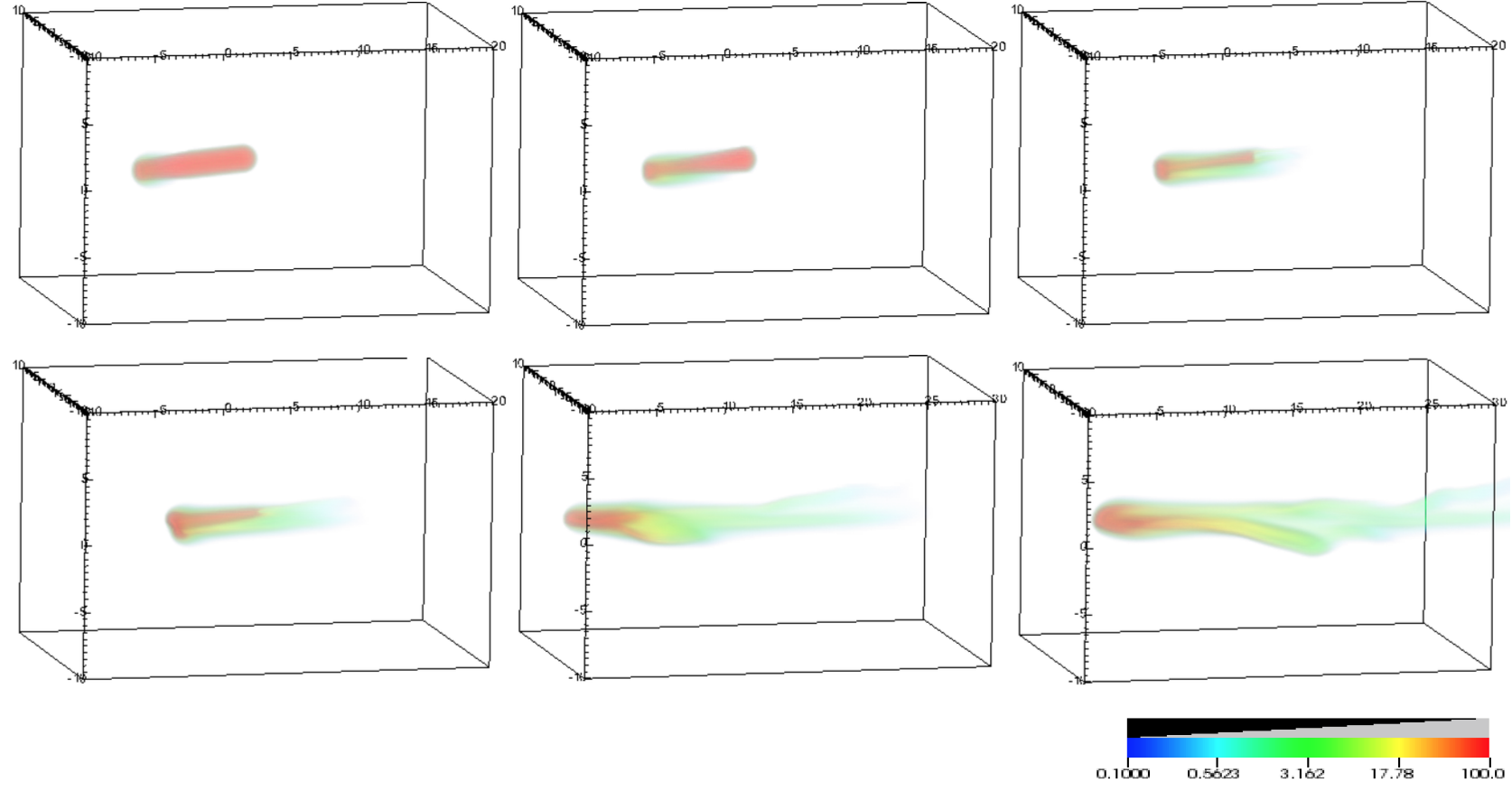} \\
\end{tabular}
\caption{A 3D volumetric rendering of model $m3c2l885$. From left to right, top to bottom the timings are $t=0.00 \,t_{\rm cs}$, $t=0.39 \,t_{\rm cs}$, $t=0.66 \,t_{\rm cs}$, $t=0.95 \,t_{\rm cs}$, $t=2.16\,t_{\rm cs}$, and $t=2.82 \,t_{\rm cs}$.}
\label{Fig11}
\end{figure*}

Figures~\ref{Fig10} and ~\ref{Fig11} show the interaction for simulation $m3c2l885$, a filament lying almost end-on to the shock front. Here, the transmitted shock travels along the entire length of the filament and a small vortex ring is visible at the upstream end. Shocks are transmitted through the sides of the filament as the transmitted and external shocks sweep through and around it. These sideways shocks, however, produce less reverberation within the filament than in the comparable filament in \citet{Pittard16a} and there is therefore considerably less voiding of the filament in the present figures. There are two main differences between this filament and that in \citet{Pittard16a}. Firstly, the bow shock is located immediately on the upstream edge of the filament and is sharply angled downstream on either side, whereas that in \citet{Pittard16a} is much more rounded and located at a distance from the filament edge. Secondly, the filament develops a smooth tail of material as the simulation progresses and broadly retains the shape of its core, unlike in the previous paper.

\subsection{Mach number dependence}
\begin{figure*}
\centering
\begin{tabular}{c}
\includegraphics[width=150mm]{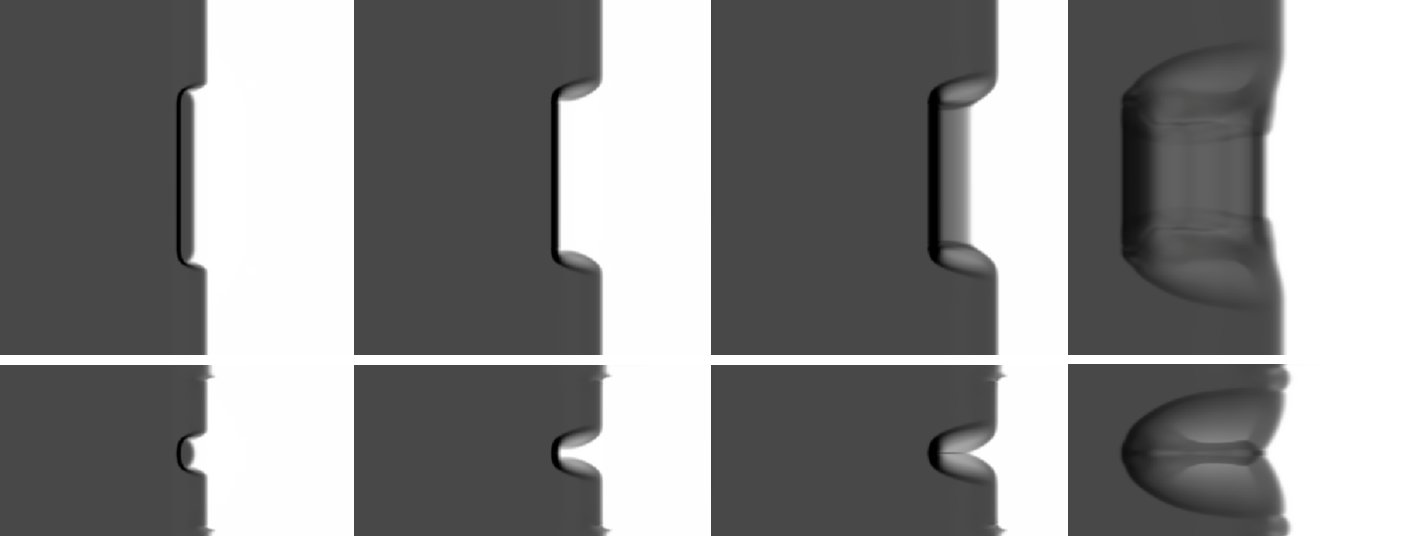} \\
\end{tabular}
\caption{The time evolution of the logarithmic density for model $m10c2l8s$ showing the $xy$ (top set of panels) and $xz$ planes (bottom set of panels). The greyscale shows the logarithm of the mass density, from white (lowest density) to black (highest density). The density scale used for this figure extends from 0 to 2.8. The evolution proceeds, left to right, with $t=0.05 \,t_{\rm cs}$, $t=0.16 \,t_{\rm cs}$, $t=0.27 \,t_{\rm cs}$, and $t=0.96 \,t_{\rm cs}$. All frames show the same region in $y$ and $z$ ($-10 < y < 10$ and $-5 < z < 5$, in units of $r_{\rm c}$). So that the motion of the cloud is clear, the first three frames show $-10 < x < 10$. The final frame shows $10 < x < 30$. In the first panel, the shock has just hit the leading edge of the filament, whereas in the second panel the shock has just passed through the filament, leaving the filament unusually compressed.}
\label{Fig13}
\end{figure*}

The Mach number dependence of the interaction is now explored. Figure~\ref{Fig13} shows the interaction of a Mach 10 shock with a filament of $\chi=10^{2}$ and a sideways orientation (simulation $m10c2l8s$). In this interaction the post-shock gas is almost as dense as the filament. It is immediately clear that the filament undergoes much greater compression in the $x$ direction compared to the same filament struck by a $M=3$ shock, and that this compression occurs over a much shorter normalised time-scale. Furthermore, the filament rapidly loses its core mass; much of the core mass has been ablated by the flow by $t\approx 1\, t_{\rm cs}$. This is a significant finding, and one which is not observed in adiabatic shock-filament interactions for an $M=10$ shock. It also contradicts the findings of \citet{Klein94} and \citet{Nakamura06} in terms of their simulations using $\gamma =1.1$ for the cloud (though it should be noted that they used a spherical cloud and not a filament). With the bow shock located so close to the upstream edge of the filament, RT fingers at each end of the filament are less in evidence, though most of the filament material is still lost from the ends of the filament. This filament can also be compared to figs. 1-3 in \citet{Pittard16a}, thus highlighting the effect of only changing $\gamma$ from 5/3 to 1.01. Compared to the adiabatic $M=10$ simulation, the quasi-isothermal shock in the current figure has a far greater density jump than the adiabatic shock and thus its interaction with the filament is much stronger. For example, it is clear that the filament in figs. 1-3 in \citet{Pittard16a} is far less compressed compared to that in model $m10c2l8s$. The quasi-isothermal filament also shows no evidence of the `three-rolled' structure present in \citet{Pittard16a}. Moreover, the transmitted shock in Fig.~\ref{Fig13} travels through the filament much more quickly than in figs. 1-3 in \cite{Pittard16a} (cf. the third panel of Fig.~\ref{Fig13} where the transmitted shock has exited the filament by approximately $t=0.16\, t_{\rm cs}$ with the fourth panel of figure 3 in \citet{Pittard16a} which shows the transmitted shock exiting the cloud at $t=0.53\, t_{\rm cs}$). Again, the bow shock is located at the upstream edge of the filament, compared to being located some distance away in figure 3 in \citet{Pittard16a}. Moreover, the quasi-isothermal filament has lost almost all its core mass by $t=1\, t_{\rm cs}$, whereas the $M=10$ adiabatic filament still has a significant amount of its core intact by this point. 

\begin{figure*}
\centering
\begin{tabular}{c}
\includegraphics[width=170mm]{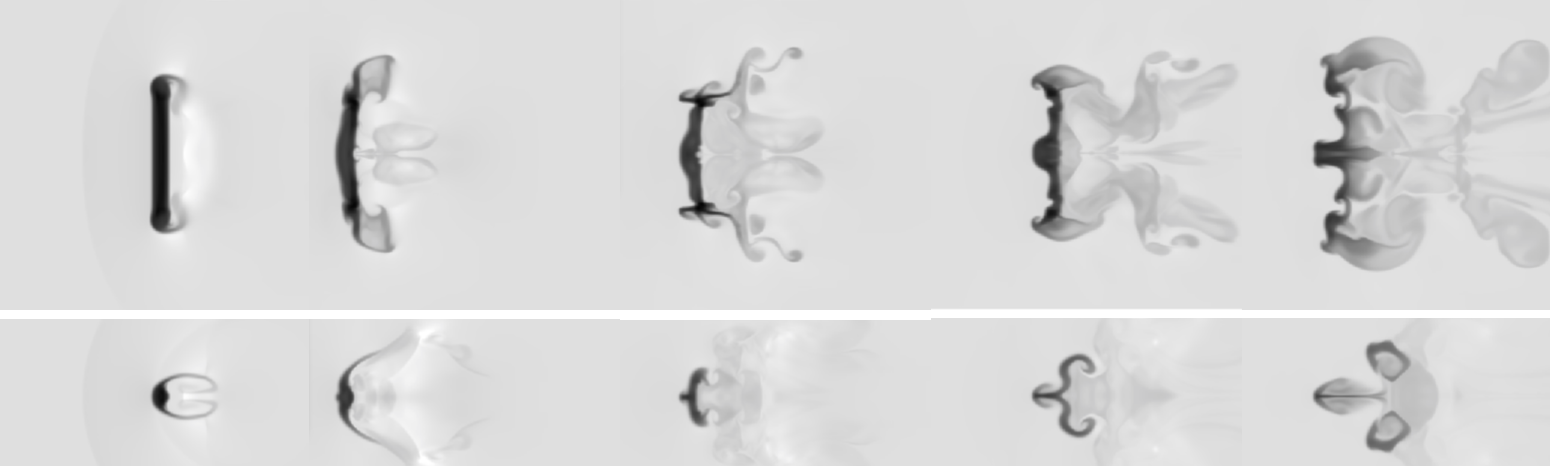} \\
\end{tabular}
\caption{The time evolution of the logarithmic density for model $m1.5c2l8s$ showing the $xy$ (top set of panels) and $xz$ planes (bottom set of panels). The greyscale shows the logarithm of the mass density, from white (lowest density) to black (highest density). The density scale used for this figure extends from 0 to 2.8. The evolution proceeds, left to right, with $t=0.58 \,t_{\rm cs}$, $t=1.92 \,t_{\rm cs}$, $t=2.75 \,t_{\rm cs}$, $t=3.58 \,t_{\rm cs}$, and $t=4.35 \,t_{\rm cs}$. All frames show the same region in $y$ and $z$ ($-10 < y < 10$ and $-5 < z < 5$, in units of $r_{\rm c}$). So that the motion of the cloud is clear, the first frame shows $-10 < x < 10$. The next three frames show $0 < x < 20$, and the final frame shows $5 < x < 25$.}
\label{Fig14}
\end{figure*}

Figure~\ref{Fig14} shows the interaction of a filament with a $M=1.5$ shock (simulation $m1.5c2l8s$). It can be seen that the interaction is more gentle than that in model $m3c2l8s$ in that filament material is not strongly stripped from the filament ends and channelled downstream behind the cloud. Instead, filament material is stripped over a longer time-scale by the flow and the filament core remains reasonably intact for much longer than in simulation $m3c2l8s$.
Many more instabilities are present on the surface of the filament throughout the simulation particularly during the early stages, compared to the filament in model $m3c2l8s$. Moreover, the RT fingers located at the filament ends are much more pronounced in the current figure and extend behind the filament rather than upstream of it. Some clumps of filament material are observed to break off the main core from around $t=2.75\, t_{\rm cs}$.

\subsection{$\chi$ dependence}
\begin{figure*}
\centering
\begin{tabular}{c}
\includegraphics[width=170mm]{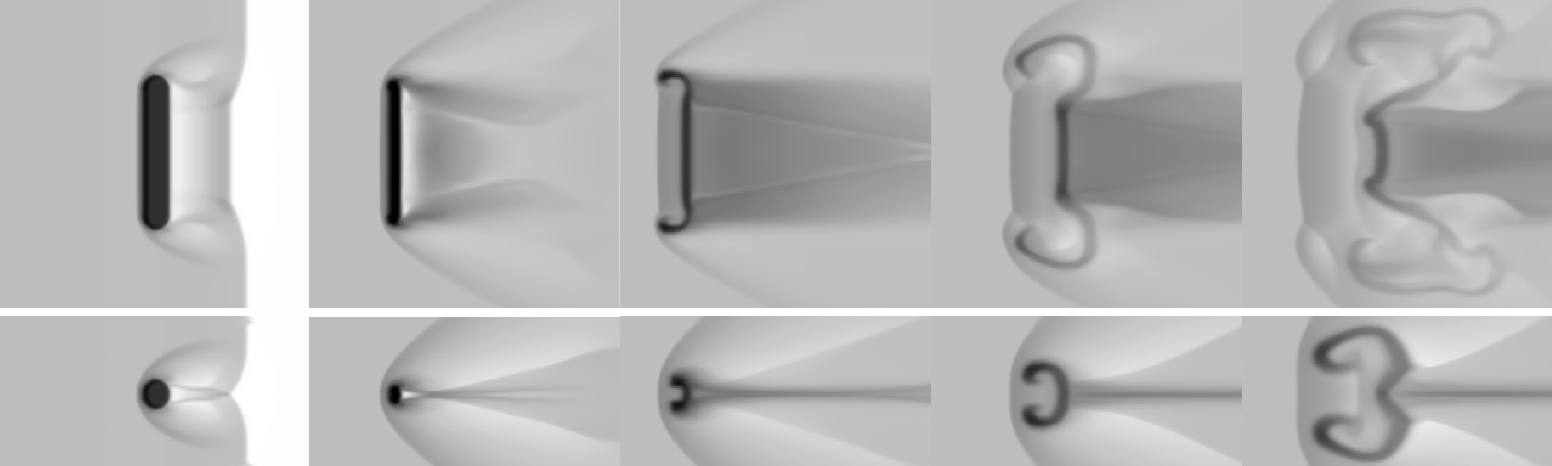} \\
\end{tabular}
\caption{The time evolution of the logarithmic density for model $m3c3l8s$ showing the $xy$ (top set of panels) and $xz$ planes (bottom set of panels). The greyscale shows the logarithm of the mass density, from white (lowest density) to black (highest density). The density scale used for this figure extends from 0 to 3.7. The evolution proceeds, left to right, with $t=0.10 \,t_{\rm cs}$, $t=0.29 \,t_{\rm cs}$, $t=0.87 \,t_{\rm cs}$, $t=1.15 \,t_{\rm cs}$, and $t=1.41\,t_{\rm cs}$. All frames show the same region in $y$ and $z$ ($-10 < y < 10$ and $-5 < z < 5$, in units of $r_{\rm c}$). So that the motion of the cloud is clear, the first frame shows $-10 < x < 10$. Frame 2 shows $-5 < x < 15$, frames 3 and 4 show $0 < x < 20$, and the final frame shows $5 < x < 25$. Note that this simulation was run at a slightly lower resolution of $R_{16}$.}
\label{Fig15}
\end{figure*}

We now investigate the dependence of the interaction on the filament density contrast. Figure~\ref{Fig15} shows the effect on the interaction when $\chi=10^{3}$. The most obvious contrast between the filament in this simulation and that in $m3c2l8s$ is the thickness of the filament once the transmitted shock has progressed through it. In addition, the filament forms a much more angular shape compared to the previous model. At later times, the filament retains its thin, `C'-shaped morphology in the $xy$ plane whilst a considerable density of filament material is present at the rear of the filament forming a long, flat and wide wake. Note that the resolution of this simulation was $R_{16}$, in contrast with the rest of the simulations.

\begin{figure*}
\centering
\begin{tabular}{c}
\includegraphics[width=170mm]{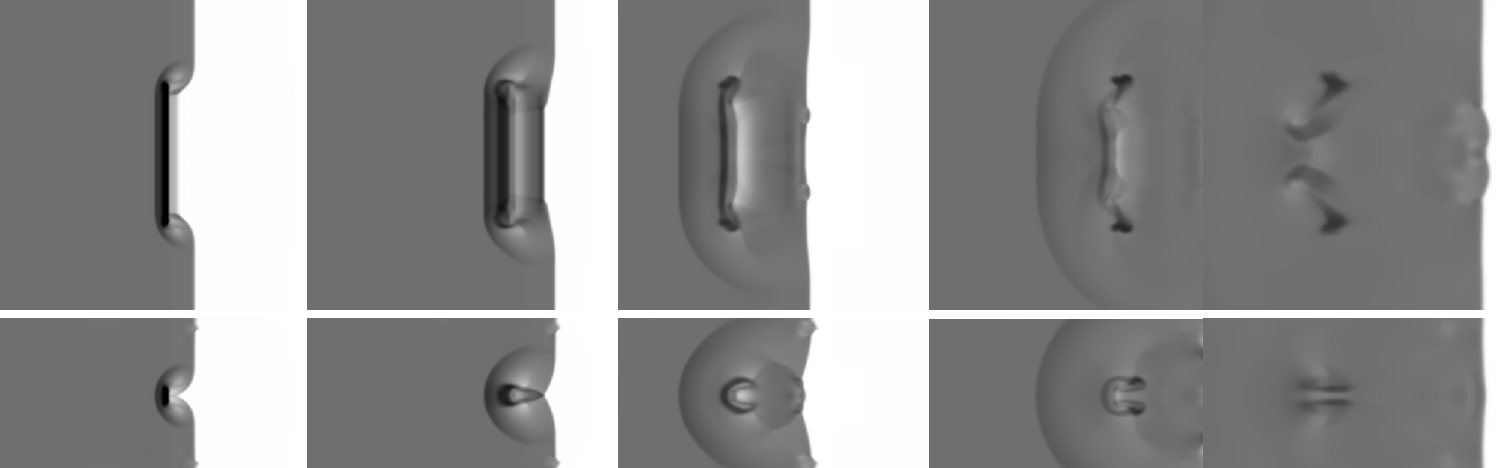} \\
\end{tabular}
\caption{The time evolution of the logarithmic density for model $m3c1l8s$ showing the $x-y$ (top set of panels) and $x-z$ planes (bottom set of panels). The greyscale shows the logarithm of the mass density, from white (lowest density) to black (highest density). The density scale used for this figure extends from 0 to 1.7. The evolution proceeds, left to right, with $t=0.33 \,t_{\rm cs}$, $t=1.05 \,t_{\rm cs}$, $t=1.90 \,t_{\rm cs}$, $t=2.95 \,t_{\rm cs}$, and $t=6.28 \,t_{\rm cs}$. All frames show the same region in $y$ and $z$ ($-10 < y < 10$ and $-5 < z < 5$, in units of $r_{\rm c}$). So that the motion of the cloud is clear, the first 2 frames show $-10 < x < 10$. Frame 3 shows $0 < x < 20$, frame 4 shows $20 < x < 40$, and the final frame shows $30 < x < 50$.}
\label{Fig16}
\end{figure*}

Figure~\ref{Fig16} shows the interaction when $\chi=10$. In this interaction the post-shock density almost exceeds that of the cloud. At $t=0.74\,t_{\rm cs}$ the filament is compressed and its filament tips are bent downstream behind the filament, unlike in similar simulations where $\chi=10^{2}$ or $10^{3}$. The bow shock is initially located very close to the filament. However, by $t=1.56\,t_{\rm cs}$ it has moved further upstream. Neither this filament nor that in Fig.~\ref{Fig15} show the three-rolled structure visible in Fig.~\ref{Fig2}. This result is interesting because the three-rolled structure was seen in the $\chi=10^{3}$ simulation presented in figure 24 in \citet{Pittard16a}.

\subsection{Statistics}
\begin{figure*} 
\centering
    \begin{tabular}{c}
\includegraphics[width=160mm]{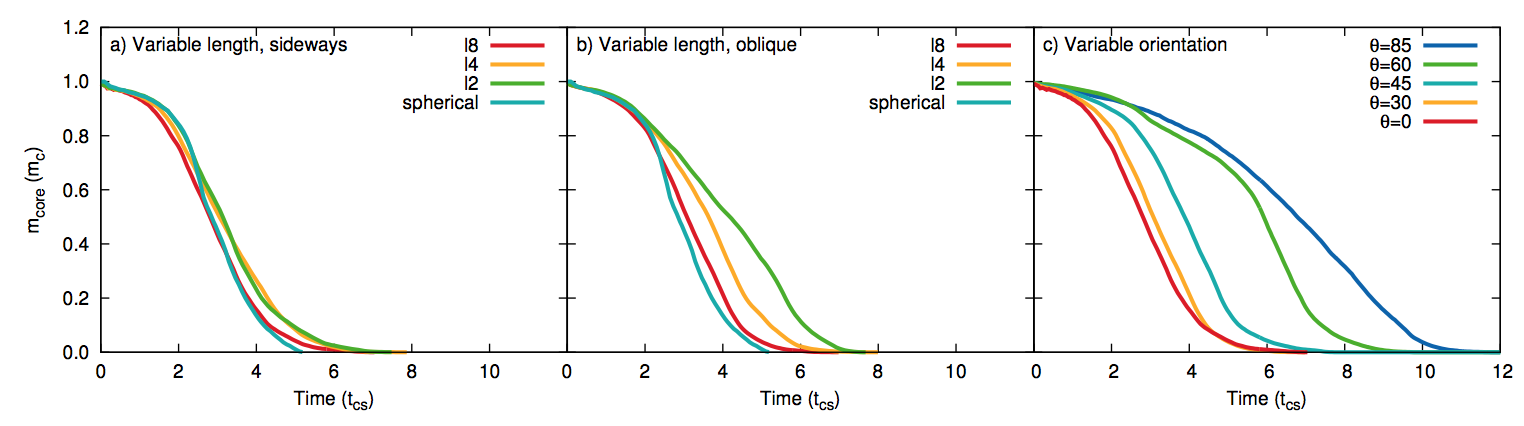} \\
\end{tabular}
  \caption{Time evolution of the core mass, $m_{\rm core}$, normalised to its initial value, for various simulations with $M = 3$ and $\chi = 10^{2}$. The left-hand panels are for `sideways' simulations, the centre panels are for simulations with $\theta = 30^{\circ}$, while the right-hand panels are for filaments with $l = 8$.}
\label{FigB}
  \end{figure*}

\begin{figure*} 
\centering
    \begin{tabular}{c}
\includegraphics[width=160mm]{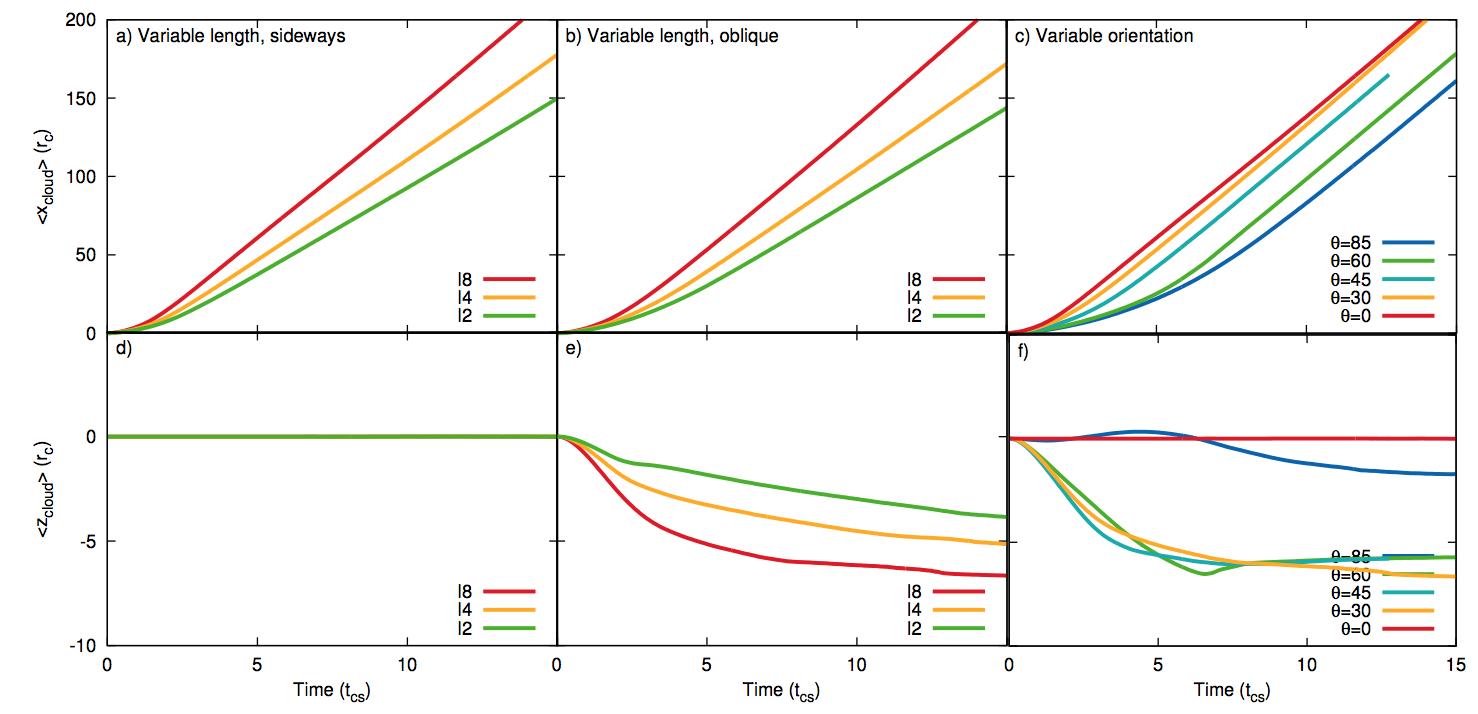} \\
\end{tabular}
  \caption{The time evolution of the $x$ and $z$ centre-of-mass position of the filament for various simulations (the same simulations as in Fig.~\ref{FigB}). The middle panels show simulations with $\theta = 30^{\circ}$, while the right-hand panels are for filaments with $l = 8$.}
\label{FigD}
  \end{figure*}

\begin{figure*} 
\centering
    \begin{tabular}{c}
\includegraphics[width=160mm]{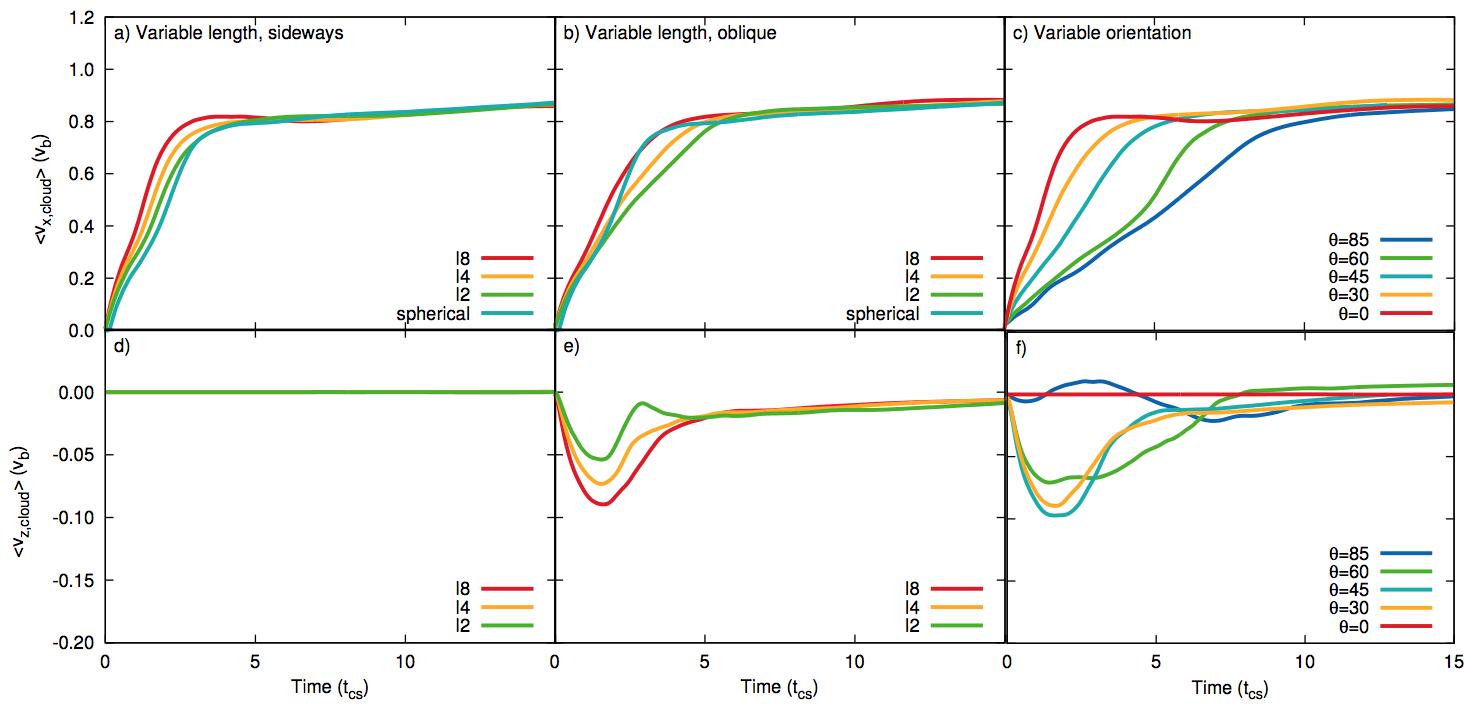} \\
\end{tabular}
  \caption{As Fig.~\ref{FigB} but showing the time evolution of the filament mean velocity in the direction of shock propagation ($\langle v_{x,\rm cloud}\rangle$) and in the $z$-direction ($\langle v_{z,\rm cloud}\rangle$). The middle panels again show simulations with $\theta = 30^{\circ}$, while the right-hand panels are for filaments with $l = 8$.}
 \label{FigF}
  \end{figure*}

\begin{figure*} 
\centering
     \begin{tabular}{c}
\includegraphics[width=160mm]{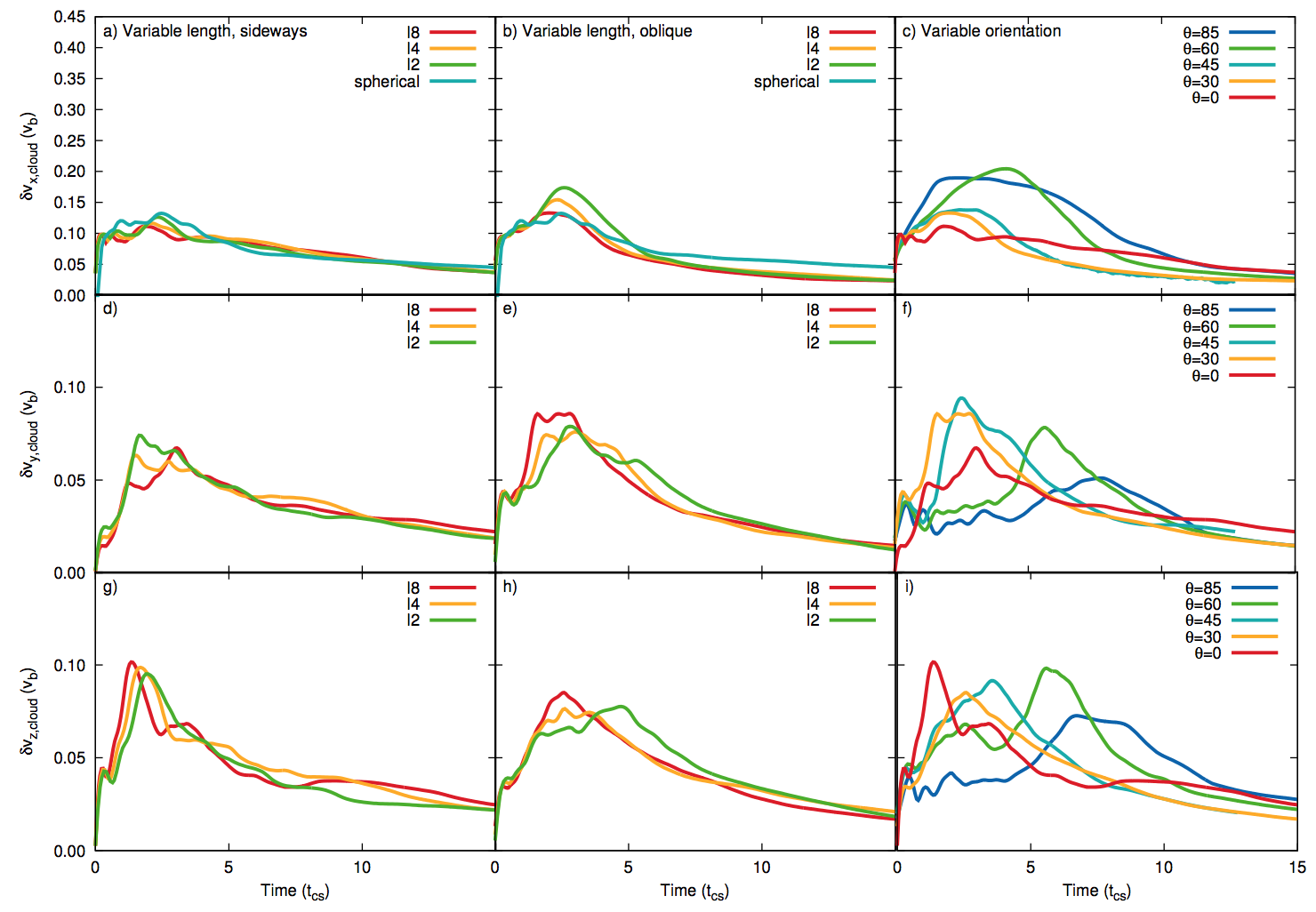} \\
\end{tabular}
  \caption{As Fig.~\ref{FigB} but showing the time evolution of the filament velocity dispersion in each direction. The middle panels again show simulations with $\theta = 30^{\circ}$, while the right-hand panels are for filaments with $l = 8$.}
 \label{FigH}
  \end{figure*}

The evolution of various global quantities of the interaction is now explored, starting with the simulations with $M=3$ and $\chi=10^{2}$. We then consider the Mach- and $\chi$-dependency of the global quantities. Figures~\ref{FigB} to ~\ref{FigH} show the time evolution of these key quantities, whilst Figs.~\ref{FigL} and ~\ref{FigM} and Table~\ref{Table1} present various time-scales taken from these simulations.

Considering first the evolution of the filament core mass, $m_{\rm core}$, Fig.~\ref{FigB} a) shows the decline in core mass for filaments of differing length with $M=3$, $\chi=10^{2}$, and a sideways orientation. Also shown are the results for a spherical cloud. It can be seen that the time taken for the core mass to be destroyed is very similar for all lengths of filament (circa $t=7\, t_{\rm cs}$), though the filament with length $l=8$ is destroyed slightly faster. Figure~\ref{FigB} b), where the filament is orientated at $\theta=30^{\circ}$ to the shock front, shows some slight variation, with shorter filaments surviving for slightly longer normalised times than longer ones, though interestingly the spherical cloud mirrors the behaviour of the filament with length $l=8$, which is odd given the greater mass of the filament compared to the spherical cloud. However, when these two figures are compared with Fig.~\ref{FigB} c), which presents filaments with various orientations but a length of $l=8$, it can be seen that there is much more variety in the rate of mass loss. Filaments aligned more closely to the shock front (i.e. filaments at $\theta=0^{\circ}$ and $30^{\circ}$) lose mass much more quickly than those oriented more `end-on' to the shock. Indeed, the filaments with very small angles of orientation have near-identical profiles, in contrast to the results presented in \citet{Pittard16a}. It is interesting to note that our results differ slightly from those in \citet{Pittard16a}, where the filament orientated at $\theta=60^{\circ}$ had the slowest degree of mass loss, in that the filament orientated at $\theta=85^{\circ}$ took the longest to be destroyed in our work. The spread in the rate of mass loss with orientation angle is also much greater than shown in figure 28 in \citet{Pittard16a} for filaments with $M=10$ and $\gamma=5/3$.
  
Figure~\ref{FigD} shows the time evolution of the $x$ and $z$ centre-of-mass positions of the filament for simulations with $M=3$ and $\chi=10^{2}$. When considering first the $x$ centre of mass the variation of the results in both panels a) and b) is immediately clear. This shows that filaments of increasing length are accelerated downstream more rapidly than shorter filaments, agreeing with the results presented in \citet{Pittard16a}. However, compared to the results in that paper, the results in Fig.~\ref{FigD} a) and b) show that the acceleration of all filaments is higher by 50 per cent. This is caused by the greater post-shock density and speed of gas in isothermal shocks. Some uniformity in filament acceleration/position of the centre of mass is found in panel c), at least in terms of filaments of orientation $\theta \leq 30^{\circ}$. However, there is also a clear gap between filaments with $\theta = 0-30^{\circ}$ and those with $\theta = 60-85^{\circ}$. This may be because the latter present less of their surface area to the shock front. This also agrees with the results of \citet{Pittard16a}. 
  
Considering the time evolution of the filament centre of mass in the $z$-direction (Figs.~\ref{FigD} d-f), no movement of the filament is observed in the $z$-direction for filaments with a sideways orientation. This is due to the effects of symmetry. When considering panel e), though, it is clear that these filaments are pushed downwards after the shock has overrun them; the filament with length $l=8$ shows far greater displacement than that with length $l=2$ because there is a greater surface area-to-volume ratio as $l$ increases. In comparison with the adiabatic simulations in \citet{Pittard16a}, where the longest filament with an orientation of $\theta=30^{\circ}$ experienced a displacement of up to $10\, r_{\rm c}$ at later times, the filaments in the present study are only displaced by between $3.5-6.5\, r_{c}$. In Fig.~\ref{FigD} f) it can be seen that the orientation of the filament has a much larger impact on the displacement of the filament in the $z$-direction. In contrast to the results presented in \citet{Pittard16a}, the filament with an orientation of $\theta=60^{\circ}$ shows as much displacement as a filament with $\theta=30^{\circ}$. However, the uplift observed in the filament orientated at $\theta=85^{\circ}$ is similar but this time occurs at around $t\approx 5\, t_{\rm cs}$ (as opposed to $t\approx 1\, t_{\rm cs}$ in the $M=10$, $\gamma =5/3$ case). This filament is also displaced much more in the $z$ direction. Filaments orientated at $\theta=30^{\circ}$ and $\theta=60^{\circ}$ experience a much greater downward motion compared to the other two filaments.
  
The time evolution of the mean filament velocity in the direction of shock propagation is shown in Fig.~\ref{FigF} a) to c). All panels show that the asymptotic velocity reached by the filaments is very similar, though there is some variation within this in that longer filaments accelerate faster than spherical clouds. The asymptotic velocity is reached by $t\approx 5 \, t_{\rm cs}$ in the first two panels a) and b). Filaments which are sideways to the shock front or orientated at $\theta =30^{\circ}$ have near identical acceleration, though again there is some variation according to filament length with shorter filaments accelerated more slowly than longer ones. Moreover, the spherical cloud is shown to accelerate faster than the filaments between $t \approx 2-3\, t_{\rm cs}$. When the filament length $l=8$ and the filament orientation is varied there is much greater variance in filament acceleration and a clear split is observed between filaments with small angles of orientation and those with large angles of orientation; the latter are accelerated up to the ambient flow velocity at a much slower rate whilst the sideways filament shows the fastest acceleration. This split is far less pronounced when $M=10$ and $\gamma =5/3$ (cf. figure 30 in \citealt{Pittard16a}).  
  
In terms of the mean velocity perpendicular to the direction of shock propagation, Fig.~\ref{FigF} d) shows no movement for filaments oriented sideways to the shock (cf. with \citet{Pittard16a} where there was slight oscillation about zero $v_{b}$). However, for obliquely-oriented filaments and filaments with $l=8$ and variable orientations there is considerable variety (panels e and f, respectively). When the filament orientation is $\theta = 30^{\circ}$ (panel e), the maximum velocity increases with increasing filament length (with the filament with $l=8$ attaining a maximum absolute velocity of almost $+0.1 v_{b}$) owing to the fragmentation of the filament core in the transverse direction. This maximum velocity soon drops back to zero again for all filaments once the filament core has been ablated by the flow and is unable to significantly fragment any further. However, in panel f) the picture is much more complex. The filament orientated at $\theta = 85^{\circ}$ shows a small net positive velocity but then oscillates between negative and zero $z$-velocities until reaching an equilibrium at zero. Meanwhile, the filament with $\theta = 30^{\circ}$ exhibits the greatest negative $z$-velocity (reaching $v_{z} \approx -0.08 v_{b}$), in agreement with the comparable filament in \citet{Pittard16a} (cf. with their figure 30). 

Figure~\ref{FigH} shows the filament velocity dispersion in each direction. Note that the vertical scale differs between each row of panels. There is reasonable agreement between the simulations when the cloud is orientated sideways or obliquely to the shock, and the maximum peak velocity distribution is almost homogenous with a peak at between 0.06 and 0.17 $v_{b}$. In contrast, panels c), f), and i) show much greater variance in the velocity dispersion in all directions and far less uniformity. The greatest maximum velocity dispersion is in $\delta v_{x, \rm cloud}$ where the filament oriented at $\theta = 60^{\circ}$ to the shock front reaches just over 0.20 $v_{b}$. Compared to figure 31 in \citet{Pittard16a}, panels f) and i) have much more variation between the models, with the filaments oriented at $\theta = 30^{\circ}$ and $\theta = 0^{\circ}$ achieving peak velocity dispersion in the $y$- and $z$-directions at earlier times. 

The orientation dependence of the interaction is shown in Figures~\ref{FigB}-\ref{FigH}. In all cases the evolution of a filament oriented at $\theta=45^{\circ}$ shows behaviour inbetween the cases for filaments oriented at $\theta=30^{\circ}$ and $60^{\circ}$. Sometimes there are large differences between the $\theta=30^{\circ}$ and $60^{\circ}$ results that the $\theta=45^{\circ}$ simulation fills in (e.g. Figures~\ref{FigB}c,~\ref{FigD}c,~\ref{FigF}c). For some other quantities the $\theta=30^{\circ}$, $45^{\circ}$ and $60^{\circ}$ results all closely agree (e.g. Figures~\ref{FigD}f and~\ref{FigF}f).

\begin{figure*} 
\centering
     \begin{tabular}{c}
\includegraphics[width=105mm]{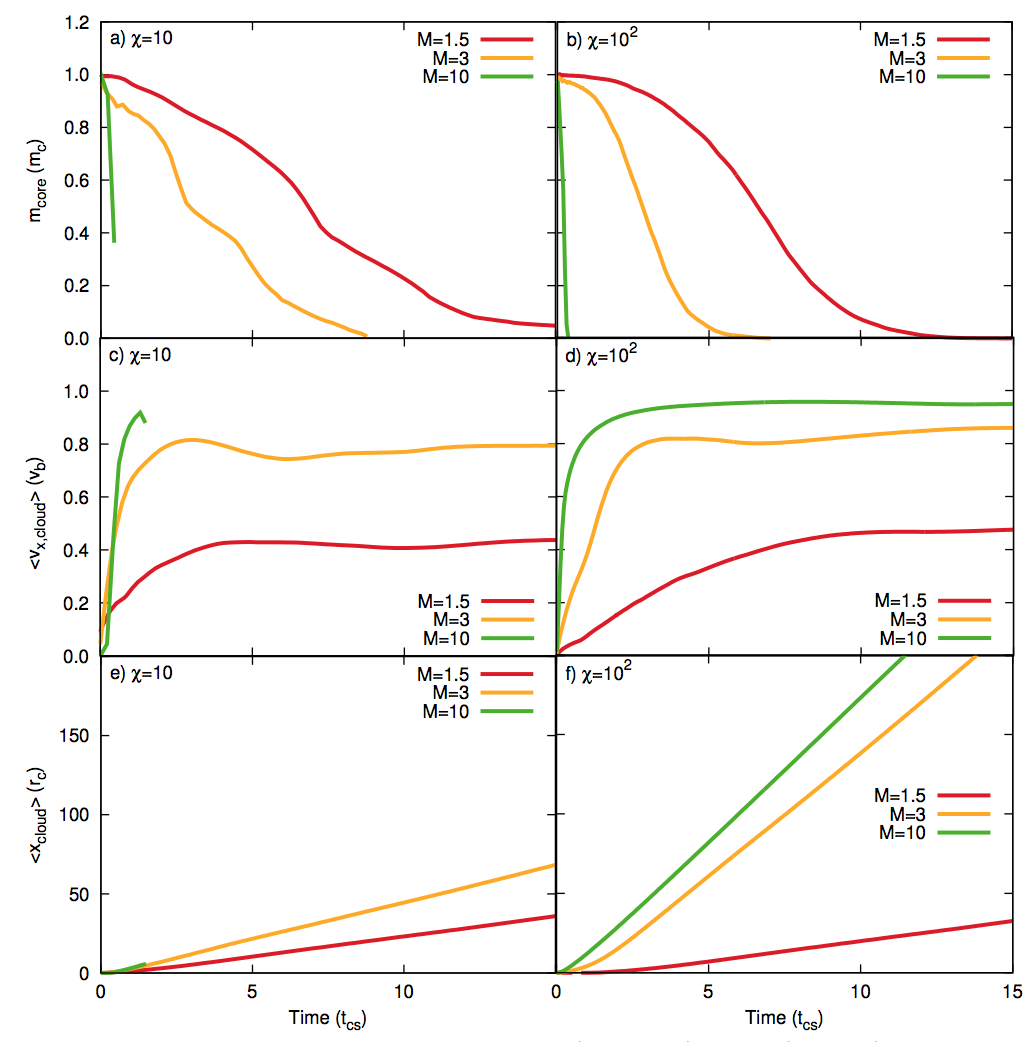} \\
\end{tabular}
  \caption{The Mach number dependence of the evolution of $m_{\rm core}$, $\langle v_{x, \rm cloud}\rangle$, and $\langle x_{\rm cloud} \rangle$, for filaments with $l = 8 \,r_{\rm c}$ and oriented sideways to the shock front.}
 \label{FigI}
  \end{figure*}

The Mach dependence of $m_{\rm core}$, $\langle v_{x, \rm cloud}\rangle$, and $\langle x_{\rm cloud} \rangle$ for filaments with $l = 8 \,r_{\rm c}$ and oriented sideways to the shock front is now investigated. Figure~\ref{FigI} shows these global quantities for simulations with $\chi = 10$ and $10^{2}$. In line with \citet{Pittard16a} we find that $m_{\rm core}$ declines more slowly as $M$ is reduced, in agreement with \citet{Pittard16a}, and that the acceleration and centre of mass position of the filament, as evidenced by $\langle v_{x, \rm cloud}\rangle$ and $\langle x_{\rm cloud} \rangle$, both reduce as $M$ decreases. Panel f) shows that the filament centre of mass is moved downstream slightly more slowly when $M=3$ compared to when $M=10$, though there is a clear difference between these two filaments and that when $M=1.5$, where the centre of mass is extremely slow to move downstream. However, the much more striking result is the extremely rapid evolution of the $M=10$ simulation (previously highlighted in Fig.~\ref{Fig13}), arising from the hugely powerful impact of a Mach 10 isothermal shock due to its extremely high post-shock density.

Note that in Fig.~\ref{FigI}, for $\chi=10$ and 100 the mean downstream velocity plots appear very similar when $M=3$, whereas the evolution of the mean downstream position appears very different. The reason for this is that the time is normalised to $t_{\rm cs}$, and $t_{\rm cs}$ is dependent on $\chi$. So while the clouds attain similar mean velocities of $\approx 0.8\,v_{\rm b}$, the denser cloud survives much longer and travels much further downstream.
  
\begin{figure*} 
\centering
   \begin{tabular}{c}
\includegraphics[width=170mm]{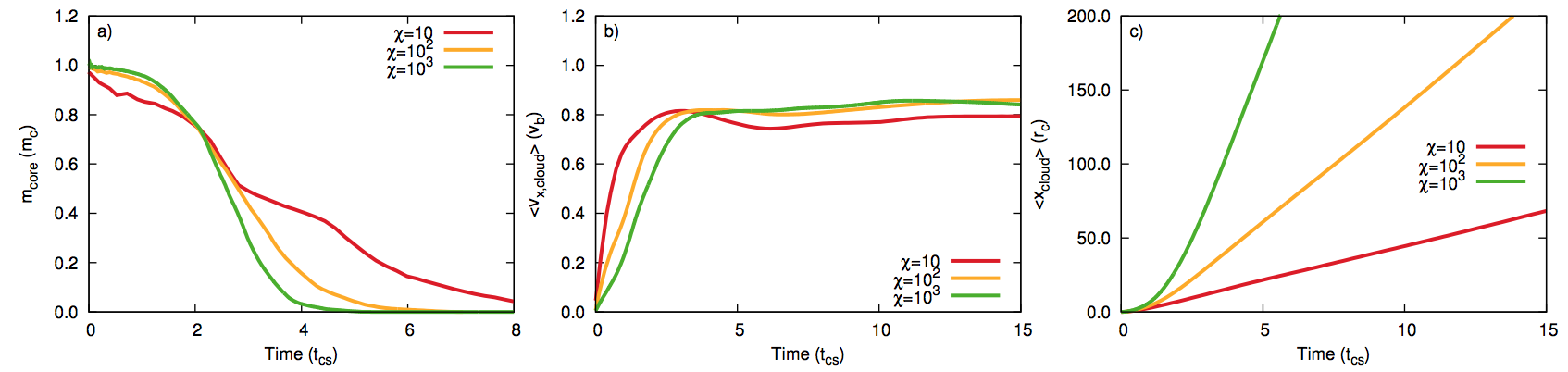} \\
\end{tabular}
  \caption{The $\chi$ dependence of the evolution of $m_{\rm core}$, $\langle v_{x, \rm cloud}\rangle$, and $\langle x_{\rm cloud} \rangle$, for $M=3$ and filaments with $l = 8 \,r_{\rm c}$ and oriented sideways to the shock front.}
 \label{FigK}
  \end{figure*}

Figure~\ref{FigK} shows the $\chi$ dependence of $m_{\rm core}$, $\langle v_{x, \rm cloud}\rangle$, and $\langle x_{\rm cloud} \rangle$ for simulations with $M=3$ and filaments with $l = 8 \,r_{\rm c}$ and oriented sideways to the shock. In terms of $\langle v_{x, \rm cloud}\rangle$ (panel b) of Fig.~\ref{FigK}) the lower $\chi$ filament experiences a faster acceleration up to the asymptotic velocity of the flow but, in comparison with the other two filaments which show very similar profiles, its velocity then drops below the asymptotic value before very slowly climbing back up again. Normalised to $t_{\rm cs}$, filaments with a lower density contrast are also slower to lose mass compared to those with a higher density contrast. Panel c) shows great variation between all filaments in the movement of the filament centre of mass downstream. When $\chi=10^{3}$ the filament experiences a rapid acceleration of its centre of mass, whereas when $\chi=10$ it is far slower to move downstream (normalised to $t_{\rm cs}$). Again, this figure compares well to figure 33 in \citet{Pittard16a}.  

 \subsubsection{Time-scales}  

\begin{figure*} 
\centering
     \begin{tabular}{c}
\includegraphics[width=170mm]{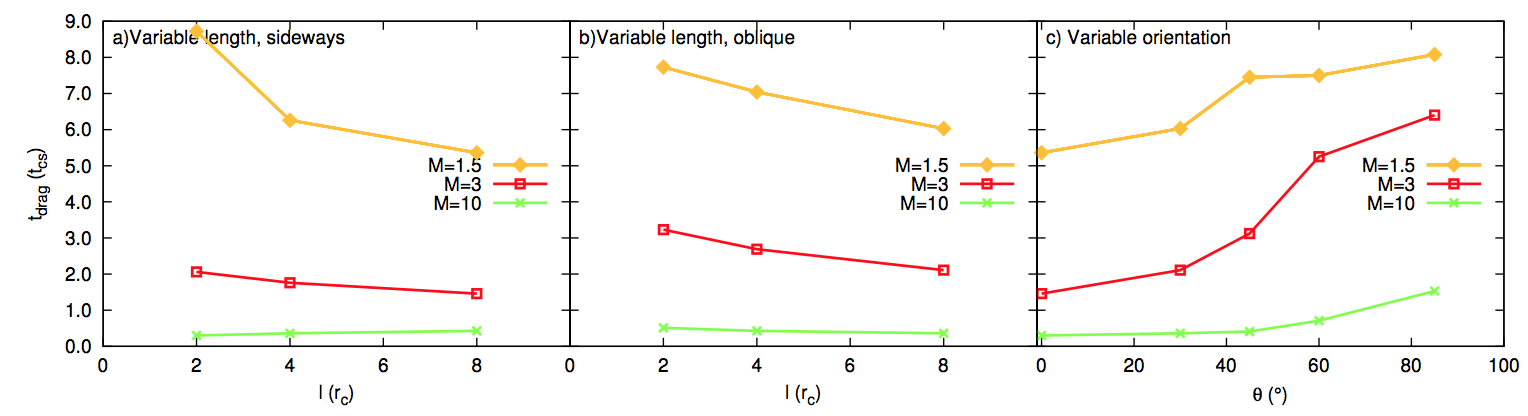} \\
\end{tabular}
  \caption{$t_{\rm drag}$ as a function of the filament length (left-hand and middle panels) and orientation (right-hand panel), from simulations with varying Mach number and $\chi = 10^{2}$. The `oblique' simulation results noted in the middle panel are for $\theta = 30^{\circ}$, while the right-hand panel is for filaments with $l = 8$.}
  \label{FigL}
  \end{figure*}   
  
 \begin{figure*} 
\centering
   \begin{tabular}{c}
\includegraphics[width=170mm]{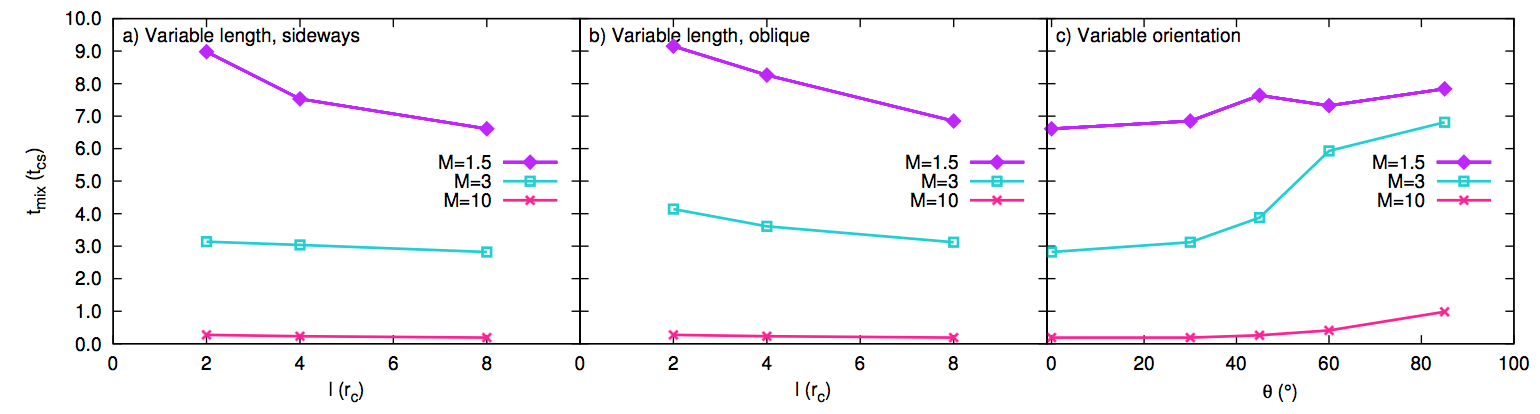} \\
\end{tabular}
  \caption{$t_{\rm mix}$ as a function of the filament length (left-hand and middle panels) and orientation (right-hand panel), from simulations with varying Mach number and $\chi = 10^{2}$. The `oblique' simulation results noted in the middle panel are for $\theta = 30^{\circ}$, while the right-hand panel is for filaments with $l = 8$.}
    \label{FigM}
  \end{figure*}   
  
Figures~\ref{FigL} and ~\ref{FigM} show values of $t_{\rm drag}$ and $t_{\rm mix}$, respectively, for simulations with $M=1.5$, $M=3$, and $M=10$ and $\chi=10^{2}$, as a function of both the filament length and orientation. Panel a) of each figure shows values taken from simulations where the filament is oriented sideways to the shock front and where the filament length is varied. Panel b) of each figure shows values taken from simulations where the filament 
is oriented obliquely (i.e. at $\theta=30^{\circ}$) and the length varied. Panel c) shows values taken from simulations where the filament has length $l=8$ but its angle of orientation is varied. Values for both time-scales (along with the filament lifetime) are also noted in Table~\ref{Table1}.

Figure~\ref{FigL} a) and b) shows that filaments in quasi-isothermal simulations with $M=3$ and $M=10$ have smaller values of $t_{\rm drag}$ compared to those in the adiabatic $M=10$ simulations presented in \citet{Pittard16a}, indicating that isothermal filaments are accelerated at a faster rate. $t_{\rm drag}$ increases with decreasing Mach number, as the interaction becomes more gentle. There is also a trend in both the adiabatic and isothermal results for $t_{\rm drag}$ to decrease as the filament length increases, though this effect is most visible in the $M=1.5$ quasi-isothermal simulations. Filaments orientated at $\theta=30^{\circ}$ to an $M=3$ shock front show slower acceleration than sideways-oriented filaments. However, the filament acceleration in simulations with $M=10$ and $M=1.5$ is comparable for sideways-on and oblique filaments with $\theta=30^{\circ}$. Panel c) shows that $t_{\rm drag}$ increases with increasing angle of orientation for simulations with $M=1.5$ and $M=3$ until $\theta=60^{\circ}$ when it begins to tail off, which is broadly in line with the adiabatic results presented in \citet{Pittard16a}. However, the results for simulations with $M=10$ do not exhibit such a tailing-off. It is clear that filaments in simulations where $M=10$ have very rapid acceleration regardless of the orientation or length of the filament (though their acceleration is less rapid as the filament becomes oriented more end-on). 

Figure~\ref{FigM} a) and b) shows a general trend for $t_{\rm mix}$ to decrease with increasing filament length for sideways and oblique filaments in simulations where $M=1.5$ and $M=3$. Sideways-orientated filaments experience mixing of core material on a slightly shorter time-scale compared to obliquely-orientated filaments. Panel c) shows that $t_{\rm mix}$ increases with the angle of orientation. There is no reduction when the filament is oriented at $\theta=85^{\circ}$ (cf. the adiabatic results in figure 34 in \citealt{Pittard16a}). The $M=3$ simulations show far more variety in mixing times compared to simulations with higher or lower Mach numbers. It is noticeable that the mixing times of filaments in simulations with $M=10$ are extremely low. Only as the angle of orientation of the filament exceeds $\theta = 30^{\circ}$ is there any increase in $t_{\rm mix}$ for these filaments.

\section{Summary and Conclusions}
In this study we investigated the HD interaction of a quasi-isothermal shock with a filament in a non-magnetised medium. This work extends the work previously presented in \citet{Pittard16a} into an isothermal regime where $\gamma =1.01$, and complements the MHD adiabatic shock-filament results presented in \citet{Goldsmith16}. We performed 3D calculations in which we varied the filament length and angle of orientation to the shock front and investigated the nature of the interaction when the shock Mach number or cloud density contrast were varied. We note the following conclusions:

i) Only sideways-oriented filaments with $\chi=10^{2}$ form a three-rolled structure, in contrast to the findings of \citet{Pittard16a}. Filaments oriented at other angles to the shock front instead form elongated structures with turbulent wakes and their morphology is dominated by the formation of a vortex ring at the upstream end of the filament. Filaments with different values of $\chi$ are bent into a `C' shape;

ii) The greater the angle of orientation, the longer and less turbulent the wake, with the filament core in model $m3c2l885$ becoming highly elongated and remaining relatively intact for some considerable time. Such filaments do not spill turbulent core material from their upper end but rather lose material smoothly from the filament sides;

iii) Most filaments showed a tendency for turbulent stripping of cloud material and the loss of clumps of material to the flow during the initial stages of the interaction. However, the filament oriented at $\theta = 85^{\circ}$ shows no such tendency towards fragmentation in the early stages (up until at least $t\approx 3\, t_{\rm cs}$), thus indicating the longer-lived nature of this filament;

iv) We find that filament length is not important for mass loss when the filament is oriented sideways. However, the orientation of the filament has a significant effect: there is a clear differentiation when the angle of orientation is varied, with those filaments oriented at angles $\theta \geq 60^{\circ}$ much slower to lose mass than those with smaller angles of orientation. This split between small and large angles of orientation is also evident in the movement of the core centre of mass in the $x$ direction and in the average velocity in the $x$ direction;

v) Values of $t_{\rm drag}/t_{\rm cs}$ for sideways and oblique filaments decline as the filament length increases. Filaments in a quasi-isothermal interaction have smaller values of $t_{\rm drag}/t_{\rm cs}$ (i.e. are accelerated faster) than filaments in an adiabatic interaction. Sideways filaments are accelerated faster than obliquely oriented ones. When the filament angle of orientation is varied, $t_{\rm drag}$ increases as the angle is increased. $t_{\rm mix}$ shows similar results to the above;

vi) The normalised evolution of the filament becomes significantly more rapid at high Mach numbers ($M=10$) due to the hugely powerful impact of a high Mach number isothermal shock.

\section*{Acknowledgements}
This work was supported by the Science \& Technology Facilities Council [Research Grants ST/L000628/1 and ST/M503599/1]. We thank S. Falle for the use of the \textsc{mg} hydrodynamics code used to calculate the simulations in this work and S. van Loo for adding SILO output to it. The calculations used in this paper were performed on the DiRAC1 Facility at Leeds, which is jointly funded by STFC, the Large Facilities Capital Fund of BIS, and the University of Leeds. The 3D volumetric renderings were created using the VisIt visualisation and data analysation software \citep{Childs12}. The data associated with this paper are available from the University of Leeds data repository (\url{https://doi.org/10.5518/573}).



\appendix

\section{Resolution Test}
\label{sec:restest}
In the shock-filament interactions presented in this paper the fluid undergoes rapid variations in time and space, and thus has turbulent-like characteristics. With flows of this nature it is important to conduct a resolution study to determine whether and at what resolution any convergence is seen. Previous work in the literature has indicated that $32-64$ cells per cloud radius is needed to capture the main flow features and for reasonable convergence of some key global quantities in interactions of a shock with a spherical cloud \citep{Pittard16b}. For shock-filament interactions, \citet{Pittard16a} found that 16 cells per filament semi-minor axis (i.e. a resolution of $R_{16}$) captures the main morphological features of the interaction, but that $R_{32}$ is the minimum needed for a more accurate description of the flow. They also found that the time evolution of $m_{\rm core}$ and $<v_{\rm x,cloud}>$ are reasonably converged by $R_{16}-R_{32}$, and that $t_{\rm mix}$ and $t_{\rm drag}$ are broadly constant at resolutions of at least $R_{8}$ (though $t_{\rm mix}$ generally declines with increasing resolution).

In the following we perform a resolution of our new isothermal shock-filament simulations.

\subsection{Time Evolution}
\begin{figure*} 
\centering
    \begin{tabular}{c}
\includegraphics[width=160mm]{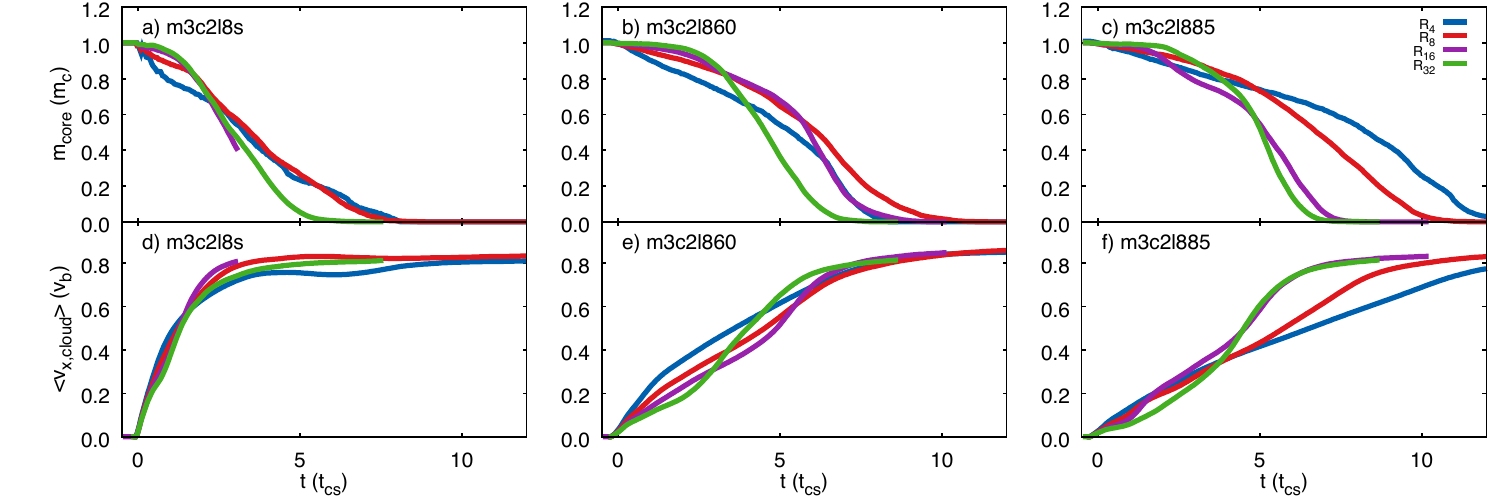} \\
\end{tabular}
  \caption{Time evolution of the core mass, $m_{\rm core}$, and the
     mean filament speed, $<v_{\rm x,cloud}>$, for
     simulations 
     (a) \emph{m3c2l8s}, (b) \emph{m3c2l860} and (c) \emph{m3c2l885}.}
 \label{FigA1}
  \end{figure*}

Fig.~\ref{FigA1} shows the time evolution of the core
mass, $m_{\rm core}$, and the mean filament speed, $<v_{\rm
  x,cloud}>$, for a number of simulations where $M=3$, $\chi=10^{2}$, and the filament length $l=8\,r_{\rm c}$. The orientation of the filament to the shock is either sideways on, or at an angle of $60^{\circ}$ or $85^{\circ}$ (the latter being nearly end-on). In all cases the $R_{4}$ and $R_{8}$ resolution simulations tend to evolve somewhat differently to the $R_{16}$ and $R_{32}$ simulations. The $R_{16}$ and $R_{32}$ simulations are reasonably coincident for the \emph{m3c2l8s} and \emph{m3c2l885} scenarios, but show greater divergence for the \emph{m3c2l860} simulation. 

\subsection{Convergence Tests}
\label{sec:restest_convergence}
To gain further insight into the effect of the grid resolution on our
simulations we examine the variation of some integral quantities
computed from the datasets at a particular moment in time. Formal
convergence demands that there is an asymptotic levelling off with
increasing resolution of a particular quantity.

\begin{figure*} 
\centering
    \begin{tabular}{c}
\includegraphics[width=130mm]{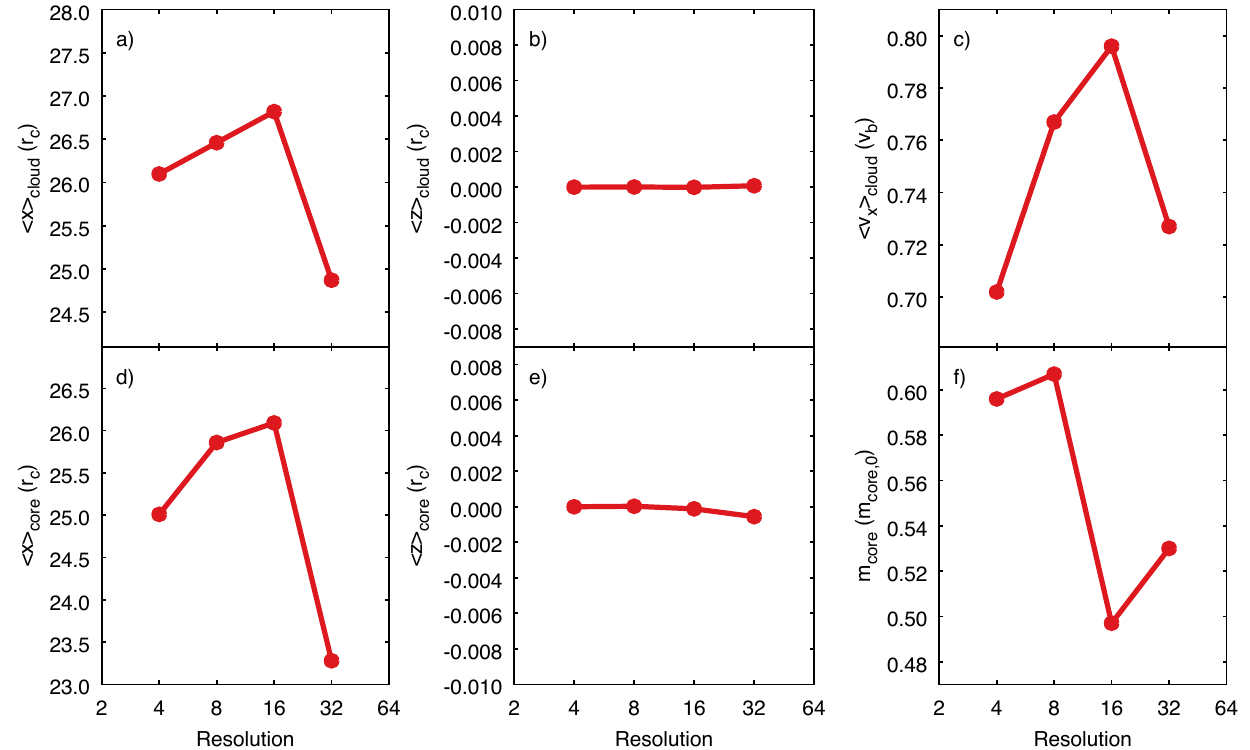} \\
\end{tabular}
  \caption{Integral quantities from simulation \emph{m3c2l8s} at $t=2.8\,t_{\rm cs}$, plotted as a function of the
     grid resolution.}
 \label{FigA2}
  \end{figure*}

The variation in $\langle x\rangle_{\rm cloud}$, $\langle
z\rangle_{\rm cloud}$, $\langle v_{\rm x}\rangle_{\rm cloud}$, $\langle x\rangle_{\rm
  core}$, $\langle z\rangle_{\rm core}$ and $m_{\rm core}$ with the
spatial resolution for simulation \emph{m3c2l8s} is shown in
Fig.~\ref{FigA2}. It is clear that there is no sign of convergence.

\begin{figure*} 
\centering
    \begin{tabular}{c}
\includegraphics[width=130mm]{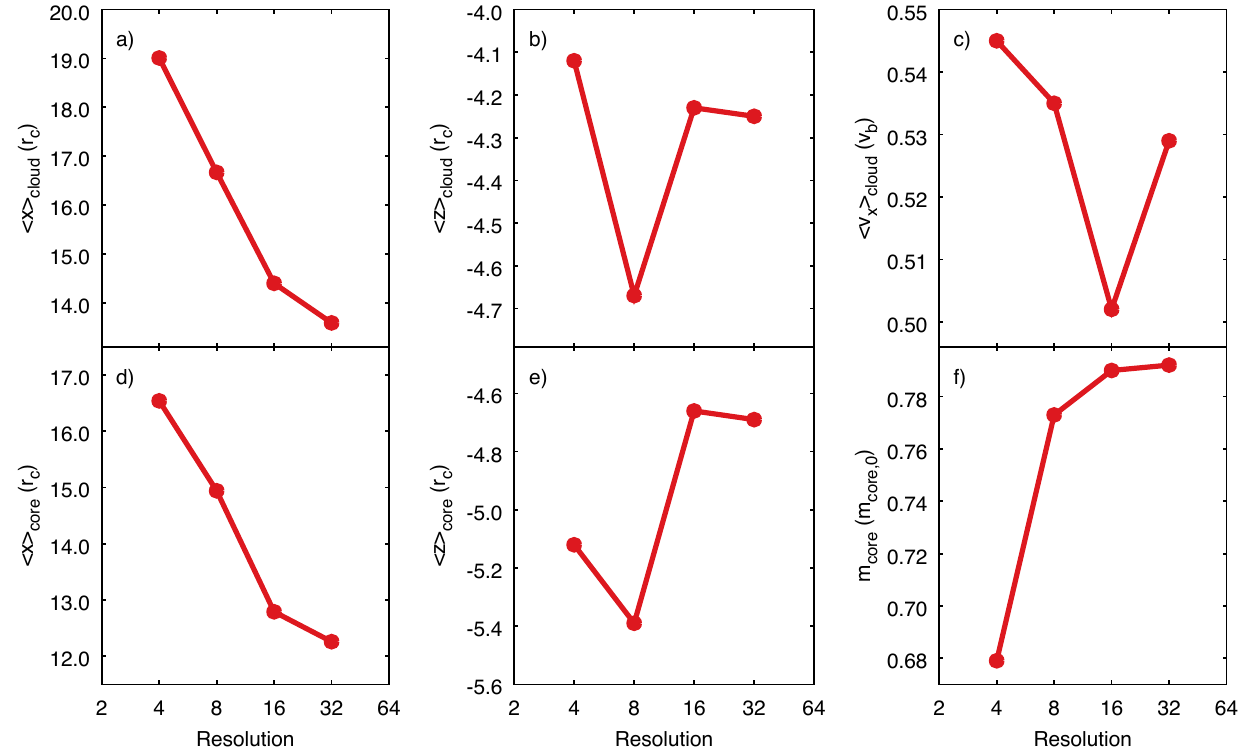} \\
\end{tabular}
  \caption{As Fig.~\ref{FigA2} but for simulation \emph{m3c2l845}.}
 \label{FigA3}
  \end{figure*}

\begin{figure*} 
\centering
    \begin{tabular}{c}
\includegraphics[width=130mm]{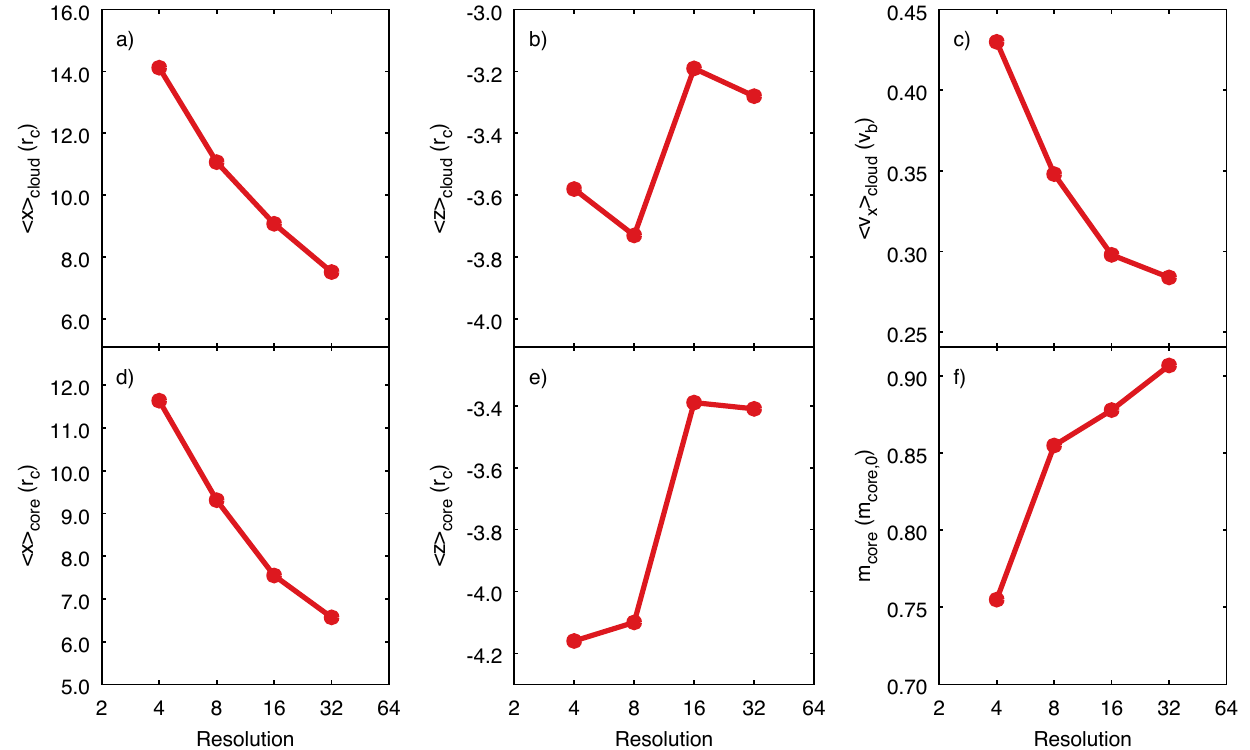} \\
\end{tabular}
  \caption{As Fig.~\ref{FigA2} but for simulation \emph{m3c2l860}.}
 \label{FigA4}
  \end{figure*}

Figs.~\ref{FigA3} and~\ref{FigA4}
examine the convergence properties for simulations \emph{m3c2l845}
and \emph{m3c2l860}. The former shows signs of convergence in $\langle x\rangle_{\rm cloud}$ and $m_{\rm core}$. The latter shows signs of convergence in $\langle x\rangle_{\rm cloud}$, $\langle v_{\rm x}\rangle_{\rm cloud}$ and $\langle x\rangle_{\rm
  core}$. However, in general the simulations do not show signs of convergence.

\subsection{Timescales}
\begin{figure*} 
\centering
    \begin{tabular}{c}
\includegraphics[width=160mm]{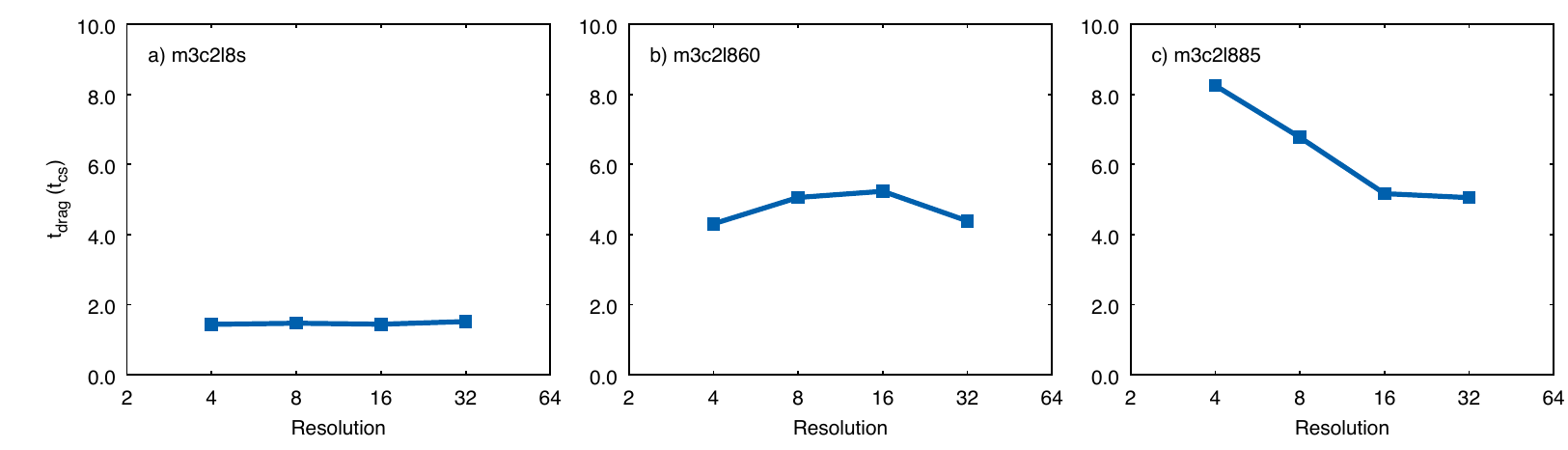} \\
\end{tabular}
  \caption{Resolution dependence of $t_{\rm drag}$ (for the
      cloud) for simulations a) \emph{m3c2l8s}; b) \emph{m3c2l860};
    c) \emph{m3c2l885}.}
 \label{FigA5}
  \end{figure*}

\begin{figure*} 
\centering
    \begin{tabular}{c}
\includegraphics[width=160mm]{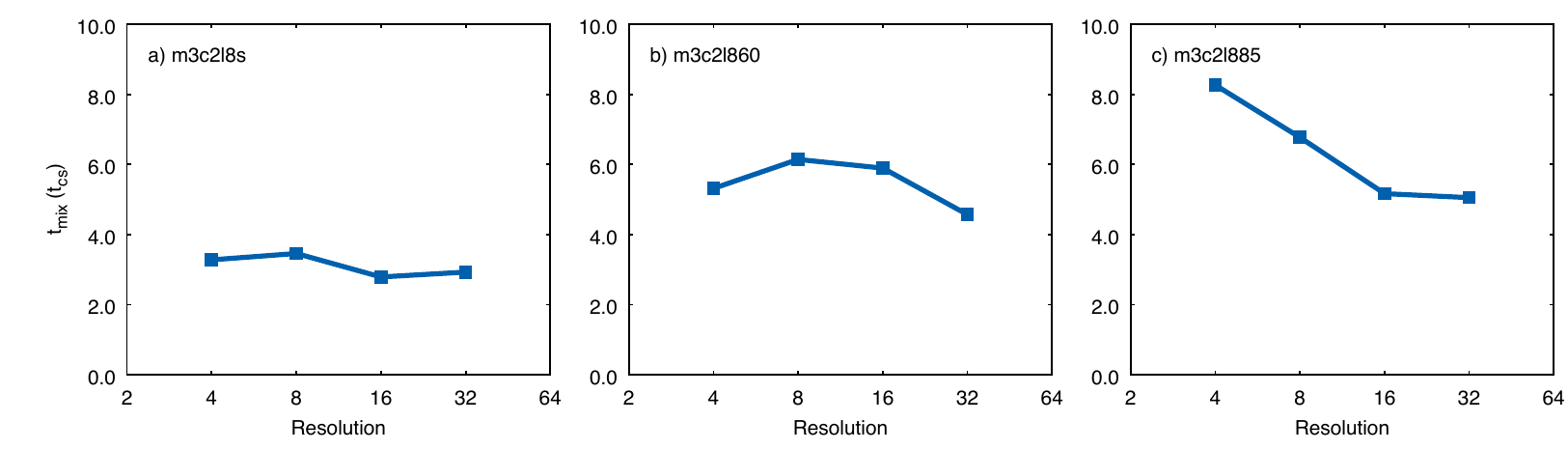} \\
\end{tabular}
  \caption{Resolution dependence of $t_{\rm mix}$ (for the
      core) for simulations a) \emph{m3c2l8s}; b) \emph{m3c2l860};
    c) \emph{m3c2l885}.}
 \label{FigA6}
  \end{figure*}

Figs.~\ref{FigA5} and~\ref{FigA6} examine the
resolution dependence of $t_{\rm drag}$ and $t_{\rm mix}$. Both quantities are broadly stable with increasing resolution for simulation \emph{m3c2l8s}. However, there are significant changes in the values of these quantities from $R_{16}$ to $R_{32}$ in simulation \emph{m3c2l860}. In contrast, in simulation \emph{m3c2l885}, $t_{\rm drag}$ and $t_{\rm mix}$ both decline significantly with increasing resolution from $R_{4}$ to $R_{16}$, but then have nearly identical values at $R_{16}$ and $R_{32}$.

Our conclusion from this study is that our simulations are generally not converged at $R_{32}$, though some properties might be close to being so. Higher resolution simulations are needed in order to further extend this study and to draw more robust conclusions.

\end{document}